\documentclass[%
prl,twocolumn,
superscriptaddress,
amsmath,amssymb,
aps,
floatfix,
]{revtex4-2}

\usepackage{graphicx}
\usepackage{dcolumn}
\usepackage{bm}
\usepackage{multirow}
\usepackage{braket}
\usepackage{amsfonts}
\usepackage{amssymb}
\usepackage{amsthm}
\usepackage{array}
\usepackage{amsmath}
\usepackage{verbatim}
\usepackage{hyperref}
\usepackage{color}
\usepackage{bbold}
\usepackage{epstopdf}
\usepackage{mathtools}
\usepackage{MnSymbol}
\usepackage{tikz}
\usepackage[export]{adjustbox}
\usepackage{subfigure}
\usepackage{threeparttable}
\usepackage{hyperref}
\usepackage{titletoc}

\renewcommand{\theequation}{\arabic{equation}}

\newcommand{\affa}{Beijing Key Laboratory of Fault-Tolerant Quantum Computing, Beijing Academy of Quantum Information Sciences, Beijing, China}
\newcommand{\affb}{Institute of Physics, Chinese Academy of Sciences, Beijing,100190, China}
\newcommand{\affc}{University of Chinese Academy of Sciences, Beijing,101408, China}
\newcommand{\affd}{School of Physics and Technology, Nanjing Normal University, Nanjing 210023, China}
\newcommand{\affe}{Institute of High Energy Physics, Chinese Academy of Sciences, Beijing, 100049, China}
\newcommand{\afff}{Hefei National Laboratory, Hefei 230088, China}

\def\ket#1{\left|#1 \right>}

\begin{document}
	
        \title{Cosmic-ray-induced correlated errors in superconducting qubit array}
        
	\author{Xuegang Li}
	\thanks{These authors contributed equally to the work.}
	\address{\affa}
	\author{Junhua Wang}
	\thanks{These authors contributed equally to the work.}
	\address{\affa}
	\author{Yao-Yao Jiang}
	\address{\affa}
	\address{\affb}
	\address{\affc}
	\author{Guang-Ming Xue}
	\address{\affa}
	\address{\afff}
	\author{Xiaoxia Cai}
	\address{\affa}
	\address{\affe}
	\author{Jun Zhou}
	\address{\affd}
	\author{Ming Gong}
	\address{\affe}
	\author{Zhao-Feng Liu}
	\address{\affe}
	\author{Shuang-Yu Zheng}
	\address{\affd}
	\author{Deng-Ke Ma}
	\address{\affd}
	\author{Mo Chen}
	\address{\affa}
	\author{Wei-Jie Sun}
	\address{\affa}
	\author{Shuang Yang}
	\address{\affa}
	\author{Fei Yan}
	\address{\affa}
	\author{Yi-Rong Jin}
	\address{\affa}
	\author{S. P. Zhao}
        \address{\affa}
        \address{\affb}
	\author{Xue-Feng Ding}
	\email{dingxf@ihep.ac.cn}
	\address{\affe}
	\author{Hai-Feng Yu}
	\email{hfyu@baqis.ac.cn}
	\address{\affa}
	\address{\afff}

	\begin{abstract}
		Correlated errors may devastate quantum error corrections that are necessary for the realization of fault-tolerant quantum computation. Recent experiments with superconducting qubits indicate that they can arise from quasiparticle (QP) bursts induced by cosmic-ray muons and $\gamma$-rays. Here, we use charge-parity jump and bit flip for monitoring QP bursts and two muon detectors in the dilution refrigerator for detecting muon events. We directly observe QP bursts leading to correlated errors that are induced solely by muons and separate the contributions of muons and $\gamma$-rays. We further investigate the dynamical process of QP burst and the impact of QP trapping on correlated errors and particle detection. The proposed method, which monitors multiqubit simultaneous charge-parity jumps, has high sensitivity to QP burst and may find applications for the detection of cosmic-ray particles, low-mass dark matter, and far-infrared photons.
	\end{abstract}
	
    \maketitle

    \bigskip
    \noindent\textbf{INTRODUCTION}
    
	Quantum bits (qubits) are inherently susceptible to various types of errors, necessitating the implementation of quantum error correction (QEC) to build logical qubits for the realization of fault-tolerant quantum computers~\cite{Gottesman_1998,aliferis2005}. The surface code is one of the promising fault-tolerance error correction schemes that leverages the topological properties of a qubit system to tolerate arbitrary local errors~\cite{Fowler_2012}. Along this direction, small-scale multiqubit correlated errors can be alleviated by the optimization of error correction methods or the allocation of more physical qubits~\cite{fowler2013, Baireuther_2018}. However, for large-scale correlated errors, the effectiveness of these strategies diminishes since the presence of non-local correlated errors can disrupt the topological properties, thereby posing significant challenges to QEC~\cite{Fowler_2014, Chubb_2021,harper2023}.
	
	For the superconducting qubits, it is found that the high-energy particles, like cosmic-ray muons and $\gamma$-rays, can significantly limit the qubit coherence times by elevating the quasiparticle (QP) density~\cite{Veps_l_inen_2020, thorbeck2022tls}. Correlated offset-charge jumps and energy relaxation errors have been observed, which are explained via numerical simulations as arising from QP bursts induced by muons and $\gamma$-rays~\cite{Wilen_2021,martinis2021}. Moreover, large-scale correlated errors have been demonstrated in Google's Sycamore superconducting processor by monitoring multiqubit simultaneous energy relaxations~\cite{McEwen_2021}, and the detection of coincident events of cosmic-ray muons and correlated errors in superconducting qubits is also reported~\cite{harrington2024synchronous}. These works show that the correlated errors can be caused by QP bursts arising mainly from muon and $\gamma$-ray radiations~\cite{Fowler2024}, and it is imperative to identify these particles and better understand their impact on the superconducting processor. 
	
\begin{figure*}[t]
    \includegraphics[width=1.0\textwidth]{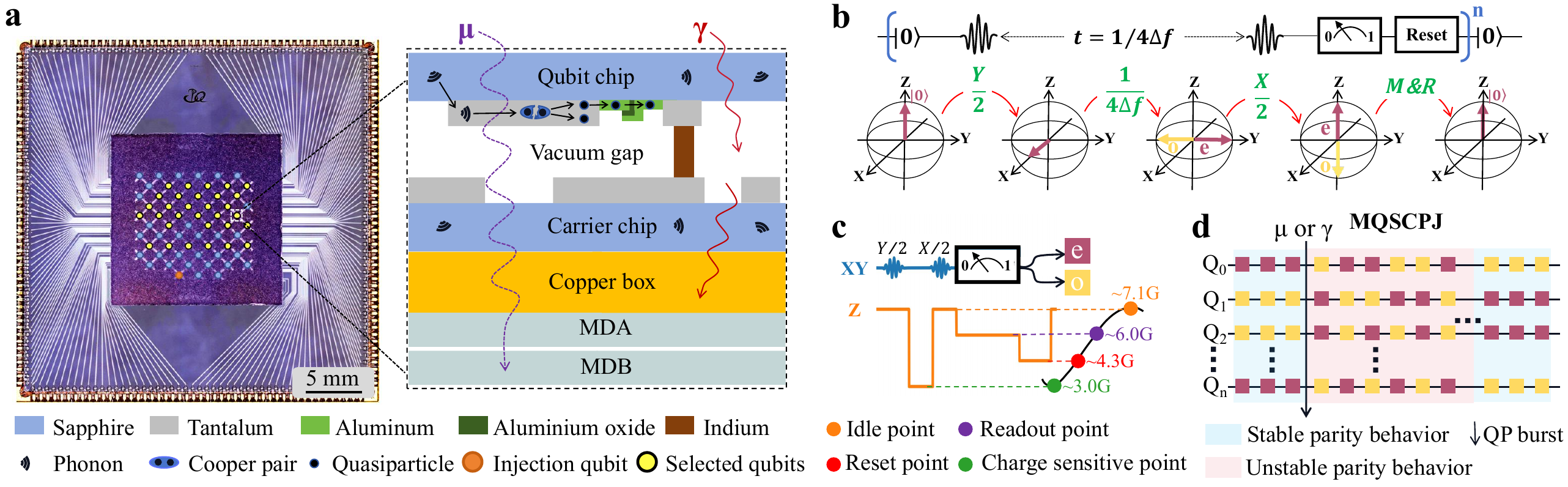}
    \caption{{\bf Device and QP burst detection.} {\bf a} Left: the optical image of the device with a bottom carrier chip and a top qubit chip containing 63 qubits and 105 couplers. The circuits on the half-transparent qubit chip are, on the back side, seen as a grid image, and color circles are added at the nodes to indicate qubits. 31 qubits in yellow are selected for measurement and 1 qubit in orange for QP injection. Right: Schematic of zoomed-in side view near a single qubit~\cite{rem}. High-energy particles, such as muons ($\mu$) and gamma-rays ($\gamma$), generate nonequilibrium phonons that break Cooper pairs thus causing QP bursts. Phonons can also be generated from the recombination of injected QPs. Two muon detectors, MDA and MDB, are located at the bottom of the copper sample box to detect muons that traverse through the device, sample box, and detectors. {\bf b} Ramsey-based sequence for measuring the charge-parity state (violetred ``e" for even state and gold ``o" for odd state) of a qubit with the corresponding evolution of the qubit state on the Bloch sphere. {\bf c} Sequence for measuring the charge-parity state using microwave control (XY), readout, and qubit frequency control (Z). The qubit frequency can be tuned to the idle point, readout point, reset point, and charge-sensitive point for qubit state control, measurement, initialization, and charge parity mapping, respectively. {\bf d} Illustration of a typical QP burst. Charge-parity states of all selected qubits (Q$\mathrm{_0}$ to Q$\mathrm{_n}$) are continuously monitored with a period of 5.6 $\mu$s. The QP bursts induced by $\mu$ or $\gamma$ are identified when there is a multiqubit simultaneous charge-parity jump (MQSCPJ).}
    \label{fig:Figure_1}
\end{figure*}

	Here, we present an experimental study of large-scale QP bursts and correlated errors induced by cosmic-ray muons and $\gamma$-rays on a 63-qubit superconducting processor (see the left panel of Fig.~\ref{fig:Figure_1}a and Methods), by continuously monitoring multiqubit correlated charge-parity jumps and bit flips, while at the same time recording the muon events via two muon detectors located beneath the sample box in the dilution refrigerator (see the right panel of Fig.~\ref{fig:Figure_1}a). We are able to separate the contributions from muons and $\gamma$-rays and observe strong coincidences between QP bursts and muon events. The occurrence time of muon-induced bit flips, estimated to be 25 minutes, remains too short to execute a QEC algorithm effectively, which typically needs operation times on the order of several hours~\cite{Xu_2022,gidney2021factor}. Moreover, we find nearly identical reduction rates of QP bursts and $\gamma$-ray events when a lead (Pb) shield is used. The percentages of QP bursts by muons and $\gamma$-rays are found to be 18.4$\%$ and 81.6$\%$, respectively.

	We also find that QPs in our case decay two to three orders of magnitude faster compared to the case in Google’s experiment~\cite{McEwen_2021}. We attribute this to the increased QPs trapped in the aluminum (Al) films of the qubit tunnel junctions. The Al films have a smaller energy gap than that of the tantalum (Ta) films which are in contact and serve as the capacitor pads, thereby forming QP traps. Our work shows that the sensitivity of charge-parity jump to QP burst is much higher than bit flip. By monitoring the charge-parity jump, using QP traps, and reducing the area of ground Ta films, the superconducting qubits could be used for the detection of cosmic-ray particles and dark matter~\cite {chou2023quantum, fink2024superconducting_1, das2024dark, hochberg2023directional} over a wide energy range.

    \bigskip
    \noindent\textbf{RESULTS}
    
    \noindent\textbf{Device and QP burst detection \label{sec:design}}
   
	Figure~\ref{fig:Figure_1}a shows the photograph of the device used in our experiment (left) and the basic process of QP bursts generated by high-energy particles (right). The device contains 63 flipmon qubits and 105 couplers~\cite{Li_2021_1, li2024_1}, which are composed of Ta capacitor pads, indium (In) bumps, and Al Josephson junctions (see also Methods and Supplementary Information~\cite{supplement}). The muons or $\gamma$-rays traverse through the sapphire substrate and generate nonequilibrium phonons, which propagate throughout the substrate transferring energy to superconducting films. During the process, Cooper pairs can be broken, thus generating excess QPs in a large scale (QP burst).  QP tunneling across the qubit's Josephson junctions, with rate linearly increasing with the QP density~\cite{martinis2009,catelani2011, Serniak_2018, Shaw2009}, leads to correlated errors. Previous works have studied the impact of high-energy particles by monitoring the energy relaxation times~\cite{Veps_l_inen_2020, Wilen_2021, McEwen_2021}, which involves bit-flip process assisted by QP tunneling~\cite{Wenner_2013}. In this work, we also use a technique of monitoring the charge-parity change when QPs tunnel. Since the charge parity will change immediately if the number of tunneling QPs are odd, this technique has a much higher sensitivity to QP burst. Theoretical estimation indicates that the probability of charge-parity change due to QP tunneling is approximately 20-80 times higher than the bit-flip probability~\cite{catelani2011}.
	
    The charge-parity state of a single transmon qubit can be measured using the Ramsey-based sequence~\cite{Rist__2013}, shown in Fig.~\ref{fig:Figure_1}b, which allows the mapping of the charge-parity state (even or odd) onto the qubit eigenstate ($\ket{0}$ or $\ket{1}$). The qubit exhibits different frequencies depending on its charge-parity state, with frequency difference 2$\Delta f$. A $\pi$ phase difference between the even and odd charge-parity states can be accumulated after an evolution time of $1/(4\Delta f)$. To accumulate a fast phase difference, it is necessary to reduce the ratio of Josephson energy ($E_J$) to charging energy ($E_C$) of the qubit~\cite{Koch_2007}. However, doing so moves the qubit away from the transmon regime, which could decrease the qubit performance.
	
    We select 31 qubits exhibiting readout fidelities above 90$\%$ for the experiment and one additional qubit for QP injection. Each qubit spectrum is engineered with a higher (~7.1 GHz) and a lower (~3 GHz) sweet spots by using asymmetrical qubit Josephson junctions, resulting in larger and smaller $E_J/E_C$ ratios. As shown in Fig.~\ref{fig:Figure_1}c, we are able to achieve high-fidelity gate operations at the higher sweet spot with $E_J/E_C\approx 83$ and $\Delta f \approx 68$ Hz (idle point) and fast charge-parity mapping at the lower sweet spot with $E_J/E_C\approx 16.5$ and $\Delta f \approx 11$ MHz (charge-sensitive point). We also achieve fast qubit reset by tuning the qubit frequency to match the readout resonator frequency (4.3 GHz) and high qubit-readout fidelity by adjusting the qubit frequency (6 GHz) to maximize readout efficiency. Overall, the Ramsey-based sequence can be repeatedly executed in 5.6 $\mu$s, enabling to continuously monitor the charge-parity state of the qubit.
	
	We explore QP bursts from the high-energy particles by monitoring the multiqubit simultaneous charge-parity jumps (MQSCPJ). As illustrated in Fig.~\ref{fig:Figure_1}d, the initial parity state of each qubit is random but stabilizes during subsequent measurements, showing a stable parity behavior. However, when a QP burst occurs, charge-parity jumps across multiple qubits are observed, showing an unstable parity behavior. After the relaxation of the QP burst, all qubits return to stable parity behavior. Note that the correlated phase error induced by charge-parity jump is negligible in transmon qubits with a large ratio of $E_J/E_C$ \cite{Koch_2007}. However, the correlated bit-flip error across multiple qubits induced by QP burst is not negligible and poses a significant obstacle to quantum error correction. Below we measure the multiqubit simultaneous bit flips (MQSBF), similar to the case in Ref.~\cite{McEwen_2021}, to evaluate correlated errors. We will show that while bit flips are closely related to correlated errors, charge-parity jumps are more sensitive to QP burst thus providing a promising method for cosmic-ray particle detection. 

\begin{figure}[t]
    \includegraphics[width=0.4\textwidth]{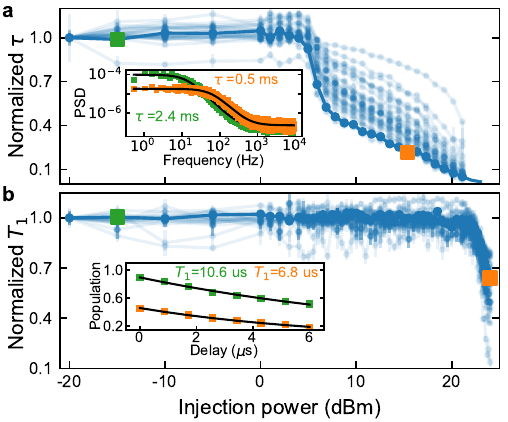}
    \caption{{\bf Response of charge-parity jump and bit flip to QP burst.} {\bf a} Average time of charge-parity jump ($\tau$) versus QP injection power for all selected qubits, normalized to the value at $-20$ dBm. Data with different darkness are for different qubits. The power spectral densities (PSD) of the charge-parity time evolution for two injection powers indicated by green and orange squares are shown in the inset, which are used to extract $\tau$. As the injection power increases to approximately 4 dBm, $\tau$ starts to decrease. {\bf b} Normalized average time of bit flip ($T_1$) versus QP injection power. The qubit energy relaxations for two injection powers indicated green and orange squares are shown in the inset, which is used to determine $T_1$. $T_1$ is seen to decrease at the injection power of approximately 22 dBm. Source data are provided as a Source Data file.}
    \label{fig:Figure_2}
\end{figure}
    
    \bigskip
    \noindent\textbf{Comparison of MQSCPJ and MQSBF\label{sec:design}}
	
	We compare the two detection methods of MQSCPJ and MQSBF. To this end, we use an injection qubit (see Fig.~\ref{fig:Figure_1}a) to produce nonequilibrium QPs~\cite{Wang_2014} (see also Methods). For each injection power, we measure the average time of charge-parity jump ($\tau$) and bit flip ($T_1$) of each qubit. As shown in the inset of Fig.~\ref{fig:Figure_2}a, $\tau$ can be determined from the power spectral density of the time evolution of charge parity obtained by continuously performing the charge-parity measurement sequence in Fig.~\ref{fig:Figure_1}c. $\tau$ is found to be 2.4 ms (green) at an injection power of -15 dBm and 0.5 ms (orange) at 15.4 dBm, respectively. Similarly in the inset of Fig.~\ref{fig:Figure_2}b, by fitting the qubit energy relaxation, $T_1$ is found to be 10.6 $\mu$s (green) and 6.8 $\mu$s (orange) at the injection powers of -15 dBm and 24 dBm, respectively. Figs.~\ref{fig:Figure_2}a and~\ref{fig:Figure_2}b display the average values of $\tau$ and $T_1$ for all selected qubits as a function of injection power, normalized to the value at the injection power of -20 dBm. We find that $T_1$ for each qubit remains unchanged until the injection power reaches approximately 22 dBm, while $\tau$ undergoes a noticeable change at an injection power of around 4 dBm. This power level, very likely, corresponds to the voltage across the qubit junctions exceeding the gap voltage, so QPs start to be generated. The injected QPs immediately cause noticeable changes of MQSCPJ but not of MQSBF.

\begin{figure*}[t]
    \includegraphics[width=0.99\textwidth]{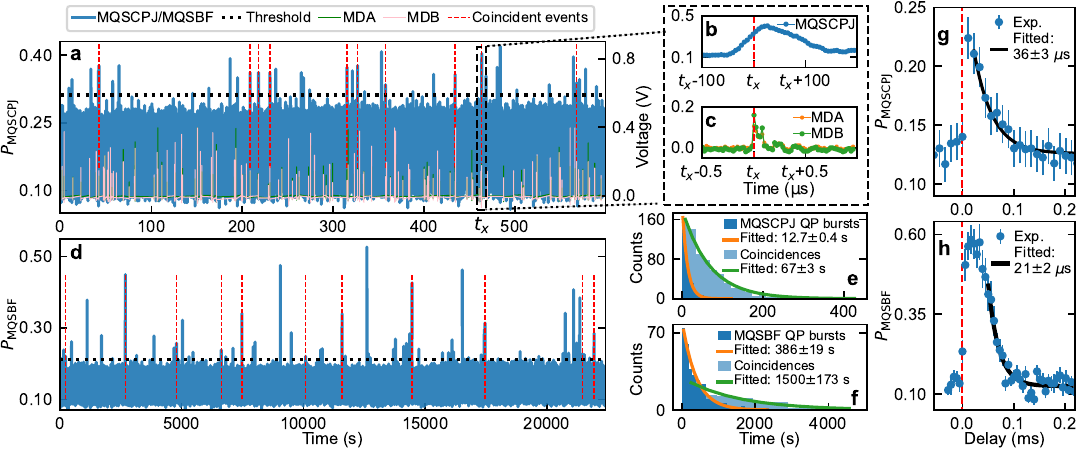}
    \caption{{\bf QP bursts by muons and $\gamma$-rays.} {\bf a} Smoothed MQSCPJ ratio $P_\mathrm{MQSCPJ}$(blue lines, left scale) in a time period of $\sim 599$ s. The horizontal dashed line indicates a threshold of 0.31 above which the MQSCPJ peaks correspond to QP bursts. Thin green and pink lines are the voltage signals of MDA and MDB (right scale). In all, 41 QP bursts and 127 muon events are detected, and 10 coincident events (marked by vertical dashed lines) are identified from the simultaneous appearance of MQSCPJ, MDA, and MDB signals as shown in {\bf b,c} near $t_x \sim 464$ s. {\bf d} Smoothed MQSBF ratio $P_\mathrm{MQSBF}$ measured over 22400 s with a threshold of 0.21. In all, 75 QP bursts, 4751 muon events (not shown due to overcrowding), and 12 coincident events are detected. {\bf e, f} Histograms of the time intervals between neighboring coincident events (light blue bars) and QP bursts (blue bars) for MQSCPJ and MQSBF experiments, respectively. Lines are the corresponding exponential fits. {\bf g, h} Averaged dynamic processes starting from each coincident events for MQSCPJ and MQSBF ratios, respectively. The error bars represent the standard deviations of the $P_\mathrm{MQSCPJ}$ and $P_\mathrm{MQSBF}$ for the coincident events. Lines are the exponential fits yielding the recombination times of 36±3 $\mu$s and 21±2 $\mu$s. To better describe the rapid process, the length of the smoothing window used in {\bf g} is 2 sampling points, while no smoothing window is used in {\bf h}. Source data are provided as a Source Data file.}
    \label{fig:Figure_3}
\end{figure*}
	
    \bigskip
    \noindent\textbf{QP bursts induced by muons \label{sec:design}}

\begin{figure}[t]
\noindent\includegraphics[width=0.45 \textwidth]{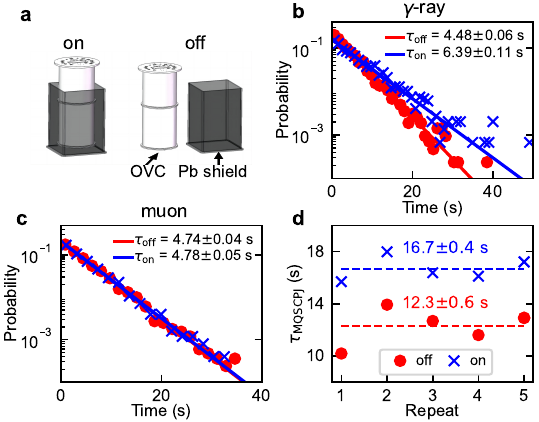}
\caption{{\bf Separation of the events induced by muons and $\gamma$-rays.} \textbf{a} Schematic of the outer vacuum chamber (OVC) of the refrigerator with (on) and without (off) Pb shield. \textbf{b,c} Probability of time intervals between neighboring events of the $\gamma$-ray detector and muon detectors with and without Pb shield. All solid lines are exponential fits to give $\tau_\mathrm{on}$ or $\tau_\mathrm{off}$. \textbf{d} The QP burst event occurrence time $\tau_\mathrm{MQSCPJ}$ is measured with and then without Pb shield, and the cycle is repeated 5 times to give an average value of 16.7$\pm$0.4 s (12.3$\pm$0.6 s) for Pb shield on (off). Source data are provided as a Source Data file.}
\label{fig:Pb_shielding}
\end{figure}
	
	We are able to observe the coincidences of QP bursts and muon events by continuously monitoring MQSCPJ or MQSBF and at the same time recording muon events by muon detectors. Employing two muon detectors (see Fig.~\ref{fig:Figure_1}a) makes it possible to distinguish signals induced by muons from those by $\gamma$-rays, as the latter is unlikely to produce simultaneous responses on both detectors, a criterion we use to identify muon events. Each muon detector has a plastic scintillator serving as the detection medium, along with a silicon photo-multiplier tube for signal collection. After the amplification at room temperature, the signal can be captured in a continuous acquisition mode by a data acquisition card.
	
	The measured MQSCPJ and MQSBF signals are found to have background noises that are mainly caused by limited qubit operation and readout fidelities. To reduce the background noise in the MQSCPJ experiment, we smooth the continuous-monitored charge-parity data for each qubit through convolution with a square window in the length of 20 sampling points. We calculate the charge-parity jumps for each qubit and average the results across all the selected qubits, which is defined as the smoothed MQSCPJ ratio (see Methods for details). In the MQSBF experiment, we calculate the ratio of the number of error qubits to the total number of selected qubits and convolve with a Gaussian window as the length of sigma about 10 sampling points, which is defined as the smoothed MQSBF ratio (see Methods for details).
	
	Figure~\ref{fig:Figure_3}a shows the typical time series of the smoothed MQSCPJ ratio for a time period up to $\sim 599$ s. Clear peaks from QP bursts above the threshold value of 0.31 corresponding to the noise level are observed. The representative time slices of a single QP burst and muon events recorded by both MDA and MDB are shown in Figs.~\ref{fig:Figure_3}b and c, respectively. The coincidence of QP burst and muon events is confirmed since the peaks fall within the time window of 100 $\mu$s. Statistically, if QP burst and muon events are completely uncorrelated, the probability of one coincidence within the duration of $\sim 599$ s is (41$\times$127)/(599 s/100 $\mu$s) $\approx$ 8.7$\times$10$^{-4}$. However, 10 coincident events are identified, as indicated by the vertical dashed lines in Fig.~\ref{fig:Figure_3}a, providing strong evidence of these QP bursts being induced by muon events. In Fig.~\ref{fig:Figure_3}a, we can see 41 QP bursts and 127 muon events. Additional QP bursts other than the coincident ones should be induced by $\gamma$-rays (see below), while additional muon events are expected considering that the detector area is larger than the area of the qubit chip. 
	
	In the MQSBF experiment, we observe 75 QP bursts and 4751 muon events over a time duration of 22400 s, as shown in Fig.~\ref{fig:Figure_3}d. A much longer time range is used since the number of QP bursts measured by MQSBF become much smaller. Because of this, Fig.~\ref{fig:Figure_3}d does not show the overcrowded detector signals of MDA and MDB. In the figure, 12 coincident events are observed as indicated by the vertical dashed lines. The uncorrelated hypothesis gives the probability of one coincidence within the duration of 22400 s is 1.6$\times$10$^{-3}$, which again confirms that these coincident QP bursts are induced by muon events. 
	
	We repeat the MQSCPJ experiments and exponentially fit the histograms of the time intervals between neighboring events to yield the average occurrence times of $12.7\pm0.4$ s for all QP bursts, $67\pm3$ s for muon-induced QP bursts, which are presented in Fig.~\ref{fig:Figure_3}e. Taking into account the size of our qubit chip (15$\times$15 mm$^2$), we can calculate a coincidence occurrence rate of 0.40$\pm$0.02 min$^{-1}$cm$^{-2}$. Since muon events affect a limited area on the qubit chip~\cite{martinez2019measurements_1}, the rate may be underestimated~\cite{supplement}. Similarly, the repeated MQSBF experiments give the average occurrence times of $386\pm19$ s for all QP bursts, $1500\pm173$ s for muon-induced QP bursts, as shown in Fig.~\ref{fig:Figure_3}f. The times are too short for QEC which typically needs operation times on the order of hours~\cite{Xu_2022,gidney2021factor}.

    \bigskip
    \noindent\textbf{QP bursts induced by $\gamma$-rays \label{sec:design}}

    In Figs.~\ref{fig:Figure_3}a and d, we see that the number of QP burst peaks exceed those in coincidence with muon detector signals. These additional peaks should be induced by $\gamma$-rays. To further see the impact of the $\gamma$-rays, we use a $\gamma$-ray detector and a Pb shield with a thickness of 1 cm that can be placed around the outer vacuum chamber (OVC) of the refrigerator to reduce the $\gamma$-ray radiation (see Fig.~\ref{fig:Pb_shielding}a). Figure~\ref{fig:Pb_shielding}b shows an average occurrence time of 4.48 s (6.39 s) without (with) the Pb shield, indicating a shielding efficiency of (1/4.48 - 1/6.39) $\times$ 4.48 = 29.9$\%$ $\pm$ 1.5$\%$. However, it shows almost no effect on the muon events with an average occurrence time of 4.74 s (4.78 s) without (with) Pb shield, as depicted in Fig.~\ref{fig:Pb_shielding}c, which corresponds to a muon flux of 0.506 $\pm$ 0.004 min$^{-1}$cm$^{-2}$ when considering the area of the plastic scintillator of 5$\times$5 cm$^2$~\cite{supplement}.

    To eliminate random fluctuations, we measure MQSCPJ and collect hundreds of events with shield off, followed by shield on, and repeat this cycle five times. As shown in Fig.~\ref{fig:Pb_shielding}d, the average values are 16.7 $\pm$ 0.4 s (12.3 $\pm$ 0.6 s) with shield on (off), indicating the average occurrence time apart from muon-induced QP bursts is 1/(1/16.7 - 1/67) s = 22.2 $\pm$ 0.8 s with shield on, and 1/(1/12.3 - 1/67) s = 15.1 $\pm$ 0.9 s with shield off. Thus, the shielding efficiency detected by MQSCPJ is (1/15.1 - 1/22.2) $\times$ 15.1 =  32$\%$ $\pm$ 5$\%$, which is close to the efficiency of 29.9$\%$ detected by the $\gamma$-ray detector. So apart from muons, QP bursts are primarily induced by $\gamma$-rays~\cite{Fowler2024}, with a ratio of approximately (1/12.3 -1/67)$\times$12.3 = 81.6$\%$ $\pm$ 1.2$\%$.

	\bigskip
    \noindent\textbf{QP trapping and particle detection \label{sec:design}}
	
\begin{figure}[t]
    \noindent\includegraphics[width=0.48 \textwidth]{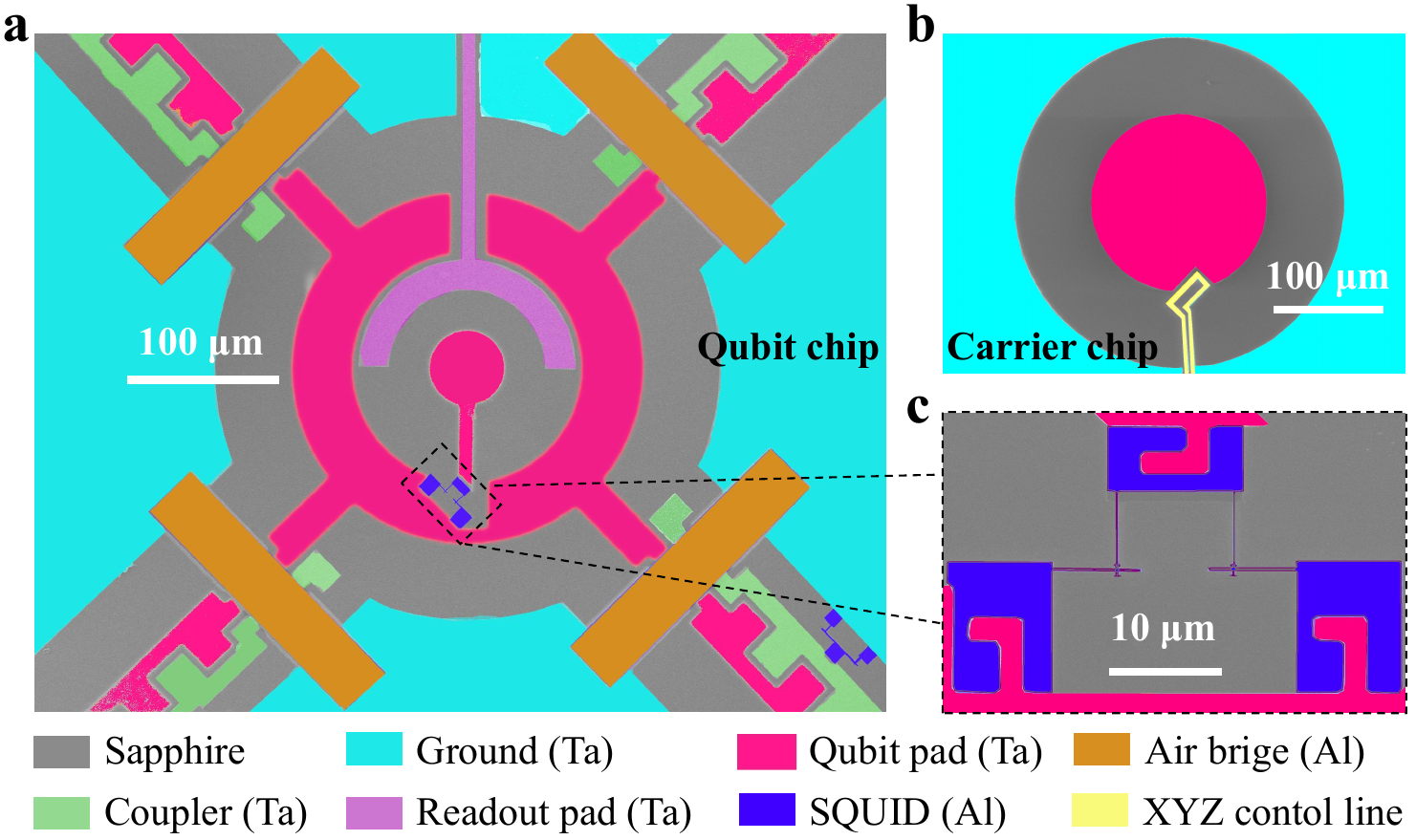}
    \caption{{\bf False-color SEM images near the center of a qubit.}  $\textbf{a}$ Qubit chip. $\textbf{b}$ Carrier chip. $\textbf{c}$ Zoomed-in view of the SQUID structure.}
    \label{fig:SEM}
\end{figure}

	To see the dynamical behavior of the QP bursts induced by muon events, we look at the processes starting from each coincident points (indicated by vertical dashed lines in Fig.~\ref{fig:Figure_3}) and average the corresponding coincident events. The results are shown in Figs.~\ref{fig:Figure_3}g and h for MQSCPJ and MQSBF experiments, respectively. We fit the tails of the averaged events and determine the recombination times to be 36$\pm$3 $\mu$s and 21$\pm$2 $\mu$s, respectively. The recombination times are roughly two to three orders of magnitude shorter than that observed in Google's Sycamore processor~\cite{McEwen_2021}, which should result largely from the increased QP density and QP trapping in the Al films of the qubit tunnel junctions. As can be seen in Fig.~\ref{fig:Figure_1}a and Fig.~\ref{fig:SEM} below in Methods, the Al films are in contact with the Ta films serving as the qubit capacitor pads. Hence, the Al films having a smaller energy gap act as QP traps that accumulate QPs. Since the QP recombination rate is proportional to the QP density~\cite{Kaplan1976, gaitskell1993non, martinis2021,McEwen_2021} which also increases the tunneling rate~\cite{Booth1987_1}, they lead to shorter recombination times (Figs.~\ref{fig:Figure_3}g and h).
	
	Specifically, the muons first deposit energy in the qubit chip (substrate), which is transferred to the Ta films as phonons near the Debye energy of $\sim 20.7$ meV. The Ta films have an energy gap of $\sim 0.69$ meV, so a number of Cooper pairs are broken leading to QP bursts. The QPs will scatter, quickly relax to the gap edge, diffuse into the Al films with a smaller gap of $\sim 0.18$ meV, and be trapped. Though the superconducting device has a complex structure with a carrier chip connected via In bumps~\cite{McEwen_2021, Li_2021_1, li2024_1}, the present situation is simpler since In bumps have a gap of $\sim 0.51$ meV and QP diffusion to the Ta pads with a larger gap can be ignored.

    The sensitivity of charge-parity jump to QP burst is much higher than that of bit flip, since the former occurs whenever odd number of QPs tunnel while the latter has a much smaller probability requiring energy conservation~\cite{Wenner_2013}. The present method using MQSCPJ may therefore be applied for the detection of cosmic-ray and dark matter particles over a wide energy range, which has recently received increasing interest~\cite {chou2023quantum, fink2024superconducting_1, das2024dark, hochberg2023directional}. With this method, the detection starts with the absorption of the energy of the particles that penetrate the qubit chip (substrate). The conversion of the energy to more QPs close to the tunnel barrier and highly detectable tunneling events are the key to high detection efficiency. Estimations considering MQSCPJ, QP trapping, and reduction of ground Ta film area show that the detection could be effective when the energy deposited in the substrate is tens of meV and above~\cite{supplement}, or for particles roughly with energies of sub-keV up to GeV. The threshold energy may vary depending on the absorption coefficient of the substrate. For dark matter particles with low coefficient~\cite{trickle2020}, the threshold is on the order of tens of keV, which is below the threshold of MeV of the presently available detectors. The method should also be applicable for the detection of far-infrared photons if they are directed to the qubit chip~\cite {fink2024superconducting_1,echternach2018single}. 

	\bigskip
    \noindent\textbf{DISCUSSION\label{sec:conclu}}

	We have provided direct evidence of QP bursts and correlated errors induced by muons and $\gamma$-rays, and successfully separated their contributions in the superconducting qubits. We find that the correlated charge-parity jumps occur more frequently than the correlated bit flips, which may also have an impact on topologically protected Majorana qubits, since they would not survive a charge-parity jump event~\cite{Albrecht_2017,Lahtinen_2017,sun2022quasiparticle}.
		
	The occurrence time of charge-parity jumps induced by muons is found to be 67 seconds, while that of bit flips is 25 minutes, which remains too high for the implementation of QEC. One effective mitigation approach is to conduct the experiment deep underground~\cite{Cardani_2021, bertoldo_2023}. However, such an approach would require high costs. Several methods propose using low-gap superconductor or normal metal as quasiparticle~\cite{Patel_2017,Bargerbos_2023} and phonon~\cite{Henriques_2019} traps, employing gap engineering\cite{mcewen2024resisting}, and suppressing the phonon transport~\cite{Iaia_2022}, in order to mitigate correlated errors. On the ground, Pb shields can be used to effectively reduce the impact of the $\gamma$-rays but not of the cosmic-ray muons. We have shown that muon detectors can operate within the refrigerator, which enables the development of muon detection arrays for the identification of the occurrence and location of muon-induced QP bursts. This can be used to build QEC circuits around the correlated error and drop a section of the device or chiplet out of the QEC protocol~\cite{Orrell_2021, Xu_2022}. 
	
	Also, the present method of monitoring multiqubit simultaneous charge-parity jumps shows high sensitivity to QP burst and has the potential for application in cosmic-ray and dark matter particle detection in the future~\cite {chou2023quantum, fink2024superconducting_1, das2024dark, hochberg2023directional}. 
	
	\bigskip
    \noindent\textbf{METHODS \label{sec:conclu}}

    \noindent\textbf{Device and Fabrication \label{sec:design}}

	Our device consists of 63 flipmon qubits and 105 couplers. The topology, design parameters, and fabrication techniques are described elsewhere~\cite{Li_2021_1, li2024_1}. As is shown in Fig.~\ref{fig:Figure_1}a, the 15$\times$15 mm$^2$ qubit chip is precisely aligned and integrated with the larger 30$\times$30 mm$^2$ carrier chip using indium bumps. The vacuum gap between these two chips is about 10 $\mu$m. The flipmon qubits feature a floating design, where one capacitor pad is on the qubit chip. The other capacitor pad is largely located on the carrier chip, with a smaller portion on the qubit chip, and the two parts are electrically connected via indium bumps. Detailed views of a flipmon qubit on the qubit and carrier chips are shown in Figs.~\ref{fig:SEM}a and b. The capacitor pads are made from 200 nm thick layers of Ta and 36 nm thick layers of Al. Two capacitor pads are electrically connected with a SQUID (superconducting quantum interference device) which is prepared using the Dolan-bridge shadow evaporation technique. The two aluminum layers used for the Dolan bridge have thicknesses of 17 nm and 19 nm, respectively. The total areas of the larger capacitor pad on the qubit chip made from Ta and Al are about 63236 $\mu$m$^2$ and 87 $\mu$m$^2$, respectively. Components around the qubit center include the couplers (green), air bridges (orange), and a readout pad (purple). The couplers mediate qubit-qubit interactions, while the air bridges ensure robust electrical connections. Fig.~\ref{fig:SEM}{\bf c} further magnifies the SQUID region (blue), showing its detailed structure. The SQUID loop enables frequency tunability of the qubit by controlling the magnetic flux. 
	
	\bigskip
    \noindent\textbf{Charge-parity jump rate vs bit-flip rate\label{sec:design}}

The charge-parity jump rate $\Gamma_{\text{e} \to \text{o}}$ divided by bit-flip rate $\Gamma_{1 \to 0}$ is given by\cite{catelani2011}:
\begin{equation}
	\small
	\frac{\Gamma_{\text{e} \to \text{o}}}{\Gamma_{1 \to 0}}=\frac{E_J(\phi_1) e^{\frac{f_{eo}(\phi_1)}{2T}} K_0\left(\frac{f_{eo}(\phi_1)}{2T}\right)
		\omega_p(\phi_2) \left(\omega_p^2(0) - \omega_p^2(\phi_1)\right)}
	{E_J(\phi_2) e^{\frac{\omega_p(\phi_2)}{2T}} K_0\left(\frac{\omega_p(\phi_2)}{2T}\right)E_C
		\left(\omega_p^2(0) + \omega_p^2(\phi_2)\right)},
\end{equation}
where $\phi_1$ and $\phi_2$ represent the normalized fluxes at which the charge-parity jump and bit-flip experiments are performed. $E_J$, $\omega_p$ and $f_{eo}$ denote the flux-dependent Josephson energy, plasma oscillation frequency, and the frequency difference between even and odd charge-parity state, respectively. $E_C$ is the charging energy. $T$ is the effective temperature of qubit. $K_0$ is the modified Bessel function of the second kind. In our experiment, we have $\phi_1 = 0.5$, $\phi_2 = 0$,  $E_J(0.5) = 4.719$ GHz, $E_J(0) = 23.8$ GHz, $\omega_p(0) = 7.379$ GHz, $\omega_p(0.5) = 3.286$ GHz, $f_{eo}(0.5) = 11$ MHz, $E_C = 0.286$ GHz, and $T \sim 40$ mK, which approximately give a ratio of $\Gamma_{e \to o}/\Gamma_{1 \to 0}$ = 18.

\bigskip
\noindent\textbf{QP injection \label{sec:design}}

As shown in Fig.\ref{fig:Figure_1}a, we select a qubit (orange) as an injection qubit to inject QPs~\cite{Wang_2014}. A microwave switch is placed at the output of a microwave source (with maximum power of 25 dBm) to generate a strong microwave pulse with the frequency set to the bare frequency of the qubit resonator ($\sim$ 4.25 GHz). With this, we generate and store photons of sufficient number, which create an oscillating voltage across the qubit Josephson junctions. When the voltage exceeds the gap voltage, a number of QPs are generated. These nonequilibrium QPs will recombine into Cooper pairs or relax to lower energy levels, producing a number of phonons that propagate throughout the substrate and break Cooper pairs leading to QP bursts, similar to the QP burst process induced by muons and $\gamma$-rays.

Generally, the intensity of the produced QP bursts increases with increasing power of the injected microwave pulse. By varying the injection power, we can control the QP burst intensity and perform the charge-parity jump and bit-flip measurements.

\bigskip
\noindent\textbf{QP burst detection by MQSCPJ and MQSBF \label{sec:design}}

The key point of detection is the ability to convert or transduce a physical process into a measurable signal, which can be processed, analyzed, and interpreted. Here, we detail how to calculate and analyze the response of MQSCPJ and MQSBF to QP bursts while taking account of system noise. Since only a few QP tunnelings occur in each qubit during a QP burst that lasts only tens of microseconds, it is challenging to distinguish a QP burst from the background noise of QP tunneling using only a single qubit. Therefore, we consider QP tunneling occurring simultaneously across multiple qubits to reliably confirm the occurrence of a QP burst. 

For the MQSCPJ experiment, we continuously monitor the charge parity of each selected qubit, recording the resulting data as a sequence of 0's and 1's. Charge parity jumps are characterized by transitions between 0 and 1. This raw data also contain random noise due to the operation and readout errors of the qubits. Therefore, convolving the raw data with a square window is well suited to suppress the noise while having little effect on the charge-parity jump behavior. The window length is set to 20 sampling points (112 $\mu s$), which is approximately three times the recombination time of QP burst in our device. This duration is also much shorter than the average interval between QP bursts (12.3 s). To further enhance the signal-to-noise ratio of the QP burst, we average the smoothed data across multiple qubits to identify the burst event, which is defined as smoothed MQSCPJ ratio ($P_\text{MQSCPJ}$):
\begin{equation}
	P_\text{MQSCPJ}(n) = \frac{1}{M} \sum_{m=1}^M \left( 0.5 - \left|\sum_{\tau=1}^L f_L(\tau) d_m(n - \tau) - 0.5 \right|\right),
\end{equation}
where $M$ is the number of selected qubits, $f_L(\tau)$ is a normalized square window function with length $L$, and $d$ is the raw data.

For the bit-flip experiments, we continuously monitor the state of each selected qubit. 
The qubit state typically remains at $\ket{1}$ but occasionally undergoes flip to $\ket{0}$ due to QP bursts. After the relaxation of the QP bursts, the qubit state returns to $\ket{1}$. We first average the raw data across multiple qubits, then considering the operation and readout errors as in the MQSCPJ experiment, we smooth the raw data by convolving with a Gaussian window with a sigma length of 10 sampling points (56 $\mu s$) to have the smoothed MQSBF ratio:
\begin{equation}
	P_\text{MQSBF}(n) = \sum_{\tau=-4L}^{4L} f_L(\tau) \left( \frac{1}{M} \sum_{m=1}^M \left( 1 - d_m(n - \tau) \right) \right),
\end{equation}
where $M$ is the number of qubits used, $f_L(\tau)$ is a normalized Gaussian window function with a sigma length $L$, and $d$ is the raw data.

The choice of detector threshold is another key point. We set threshold values conservatively in the MQSCPJ and MQSBF experiments, which almost eliminates the concerns about QP bursts caused by long-term random fluctuations. The thresholds are set to 6 (6.6) sigma deviation of the mean value of the smoothed MQSCPJ (MQSBF) ratio in the duration of $\sim$ 599 s (22400 s).

\bigskip
\noindent\textbf{DATA AVAILABILITY}

Source data are provided with this paper. The source data files are available at \href{https://doi.org/10.6084/m9.figshare.28815041}{https://doi.org/10.6084/m9.figshare.28815041}. Other data are available from the corresponding author upon request.


\bigskip
\noindent\textbf{ACKNOWLEDGMENTS}

We appreciate the helpful discussion with T. Xiang. We thank X. Liang, C. Yang, C. Wang, T. Su, C. Li, and Z. Mi for their contribution to the quantum chip preparation. We acknowledge supports from the National Natural Science Foundation of China (Grant Nos. 92365206, 11890704, 12322413, 92476206, 12104055,  22303005, 12275133, and E311455C), Innovation Program for Quantum Science and Technology (No.2021ZD0301802), and Chinese Academy of Sciences Laboratory Innovation Fund E3291TD. M.G. and Z.-F. L. acknowledge the support from the Innovative Projects of Science and Technology at IHEP, CAS.

\bigskip
\noindent\textbf{AUTHOR CONTRIBUTIONS STATEMENT}

X.L., J.W. and H.-F.Y. conceived the experiment, X.L., G.-M.X., M.C. and W.-J.S. designed and fabricated the quantum device, X.L., J.W. and Y.-Y.J. performed the measurement, S.-Y.Z., D.-K.M. and J.Z. simulated the propagation of phonons in the substrate, M.G., Z.-F.L., and X.-F.D. calculated and verified the relevant data of cosmic rays, X.L., J.W., S.Y., X.C., F.Y., Y.-R.J., and S.P.Z. analyzed the experimental data, X.L., J.W., S.P.Z., and H.-F.Y. wrote the manuscript. All authors discussed the results and the manuscript.

\bigskip
\noindent\textbf{COMPETING INTERESTS}

The authors declare no competing interests.


\bigskip
\noindent\textbf{REFERENCES}

%

\clearpage

\onecolumngrid
\section{Supplementary information for cosmic-ray-induced correlated errors in superconducting qubit array}
\vspace{1cm}

\titlecontents{section}[50pt] 
  {\addvspace{4ex}\bfseries} 
  {\contentslabel[\thecontentslabel]{2em}} 
  {} 
  {\titlerule*[1pc]{.}\contentspage} 

\titlecontents{subsection}[70pt] 
  {\addvspace{0.5ex}} 
  {\contentslabel[\thecontentslabel]{2em}} 
  {} 
  {\titlerule*[1pc]{.}\contentspage} 

\titlecontents{subsubsection}[90pt] 
  {\addvspace{0.5ex}} 
  {\contentslabel[\thecontentslabel]{2em}} 
  {} 
  {\titlerule*[1pc]{.}\contentspage} 

\startcontents
\printcontents{}{0}{}


\clearpage

\twocolumngrid

\vspace{1cm}

\setcounter{figure}{0}
\renewcommand{\figurename}{Figure}
\renewcommand{\thefigure}{S\arabic{figure}}
\renewcommand*{\theHfigure}{\thefigure}

\setcounter{table}{0}
\renewcommand{\thetable}{S\arabic{table}}
\renewcommand*{\theHtable}{\thetable}

\setcounter{equation}{0}
\renewcommand{\theequation}{S\arabic{equation}}
\renewcommand*{\theHequation}{\theequation}

\section{Device and experimental setup}

\subsection{Device}

\begin{figure}[b]
\centering
\includegraphics[width=0.35 \textwidth]{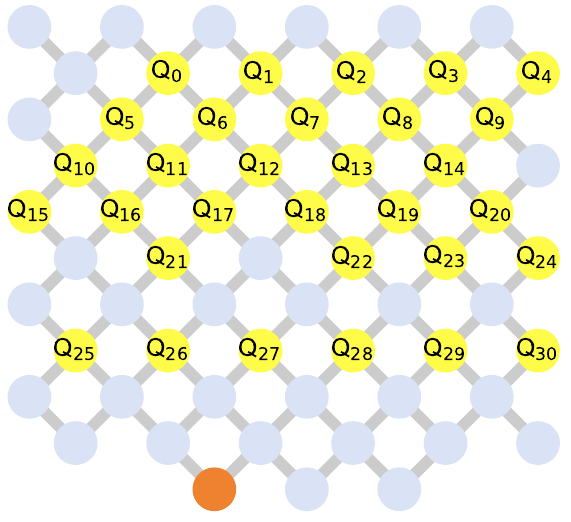}
\caption{{\bf Schematic of the superconducting qubit array used in the present experiment.} The device contains 63 flipmon qubits (circles) and 105 couplers (bars). We select 31 qubits (yellow circles) in the experiment together with one qubit (orange circle) for injecting quasiparticles.}
\label{fig:qubits layout}
\end{figure}

\begin{table*}[h]
\centering
\caption{Summary of the device and material parameters.}
\label{tab:all parameters}
\centering
\begin{tabular}{p{11cm}|c|c}
\hline
Parameter & \multicolumn{2}{c}{Value}\\
\hline
Size of qubit chip (length$\times$width$\times$thickness in mm$^3$) & \multicolumn{2}{c}{15$\times$15$\times$0.43} \\ 
\hline
Size of carrier chip (length$\times$width$\times$thickness in mm$^3$) & \multicolumn{2}{c}{30$\times$30$\times$0.43} \\ 
\hline
Thickness of Al film (nm) & \multicolumn{2}{c}{36} \\ 
\hline
Thickness of Ta film (nm) & \multicolumn{2}{c}{200} \\ 
\hline
Al film area of the capacitor pad on qubit chip ($\mu$m$^2$) & \multicolumn{2}{c}{87} \\ 
\hline
Ta film area of the capacitor pad on qubit chip ($\mu$m$^2$) & \multicolumn{2}{c}{63236} \\ 
\hline
Area of the Josephson junction ($\mu$m$^2$) & \multicolumn{2}{c}{0.029} \\ 
\hline
Superconducting gap of Al, $\Delta_\text{Al}$ ($\mu$eV) & \multicolumn{2}{c}{180} \\ 
\hline
Superconducting gap of Ta, $\Delta_\text{Ta}$ ($\mu$eV) & \multicolumn{2}{c}{690} \\ 
\hline
Superconducting gap of In, $\Delta_\text{In}$ ($\mu$eV) & \multicolumn{2}{c}{510} \\ 
\hline
Cooper pair density of Al, $n_\text{cp}$ (m$^{-3}$) & \multicolumn{2}{c}{4.18$\times$10$^{24}$} \\ 
\hline
Average $T_1$ at low sweet point ($\mu$s) & \multicolumn{2}{c}{33} \\
\hline
Average $T_2^*$ at low sweet point ($\mu$s) & \multicolumn{2}{c}{4.2}  \\
\hline
Average $T_1$ at high sweet point ($\mu$s) & \multicolumn{2}{c}{9.1} \\
\hline
Average $T_2^*$ at high sweet point ($\mu$s) & \multicolumn{2}{c}{4.7}  \\
\hline
Average qubit frequency at lower sweet point (GHz) & \multicolumn{2}{c}{2.97} \\
\hline
Average qubit frequency at higher sweet point (GHz) & \multicolumn{2}{c}{7.13}  \\
\hline
Average decay rate of readout resonator $\kappa$ (MHz) & \multicolumn{2}{c}{2.7} \\
\hline
Average cavity frequency (GHz) & \multicolumn{2}{c}{4.33}  \\ 
\hline
Average $E_\text{J}$ at higher sweet point (GHz) & \multicolumn{2}{c}{23.8} \\ 
\hline
Average $E_\text{J}$ at lower sweet point (GHz) & \multicolumn{2}{c}{4.719}\\  
\hline
Average $E_\text{C}$ (MHz) & \multicolumn{2}{c}{286} \\
\hline
Average $E_\text{J}$/$E_\text{C}$ at higher sweet point & \multicolumn{2}{c}{83.2} \\ 
\hline
Average $E_\text{J}$/$E_\text{C}$ at lower sweet point & \multicolumn{2}{c}{16.5} \\ 
\hline
Average readout fidelity of ground state & \multicolumn{2}{c}{0.97} \\
\hline
Average readout fidelity of excited state & \multicolumn{2}{c}{0.92}  \\
\hline
Average frequency difference between even and odd state $\Delta f_\text{oe}$ (MHz)& \multicolumn{2}{c}{11} \\
\hline
\end{tabular}
\end{table*}

As shown in Figure~\ref{fig:qubits layout}, our device includes 63 flipmon qubits (circles) and 105 couplers (bars). The device structure and fabrication are discussed in the main text and are described in detail elsewhere~\cite{Li_2021, li2024}. We selected 31 qubits (readout fidelity above 0.9) for the experiment. These qubits are labeled from Q$_0$ to Q$_{30}$ (yellow circles) in Figure~\ref{fig:qubits layout}. The area covered by these qubits is about 4.2 $\times$ 7.8 mm$^2$, where the distance of nearest neighbor qubits is 1 mm. The injection qubit (orange circle) is chosen as far away from the nearest selected qubit as possible. The experimental and simulation parameters are listed in detail in Table~\ref{tab:all parameters}.

\subsection{Measurement setup}

\begin{figure*}[t]
\centering
\includegraphics[width=0.6 \textwidth]{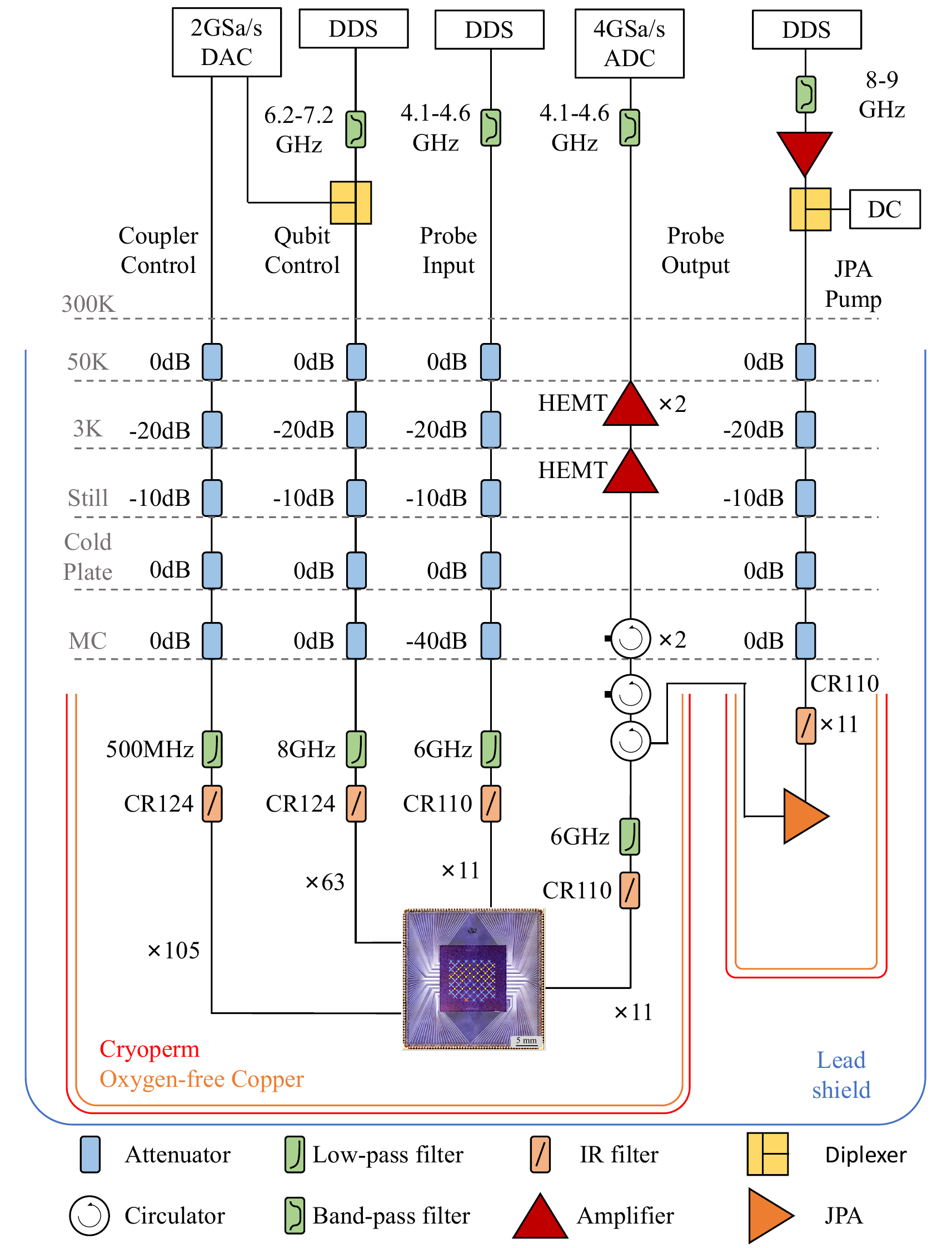}
\caption{{\bf Schematic of the measurement setup.}}
\label{fig:Measurement setup}
\end{figure*}

Our measurement setup is shown in Figure~\ref{fig:Measurement setup}. Each qubit is controlled by independent XY and Z signals, which are generated by combining the two signals at room temperature using a duplexer. The XY signal controls the qubit XY rotation, while the Z signal controls the qubit frequency. The duplexer allows the two signals to be transmitted over the same control line. An infrared (IR) filter (Eccosorb$^{\circledR}$ CR124) is used to attenuate the XY control signal by approximately 30\,dB, which has negligible effect on the Z control signal. The Z control signals used for qubits as well as couplers are generated by a 2 GSa/s digital-to-analog converter (DAC). The XY control signals, the probe input signals, and the pump signals of the Josephson parametric amplifier (JPA) are all generated by direct digital synthesis (DDS) with a sampling rate of 6 GSa/s. The probe output signals are directly captured by an analog-to-digital converter (DAC) with a sampling rate of 4 GSa/s. Since the probe input and output signals are generated by the RFSoC (Radio Frequency System-on-Chip) where the FPGA (Field-Programmable Gate Array) combines with high-speed ADCs and DACs in a single chip, stable phases can be guaranteed for each experiment. The utilization of DDS to directly generate the pump signals of the JPA instead of using the microwave/RF signal generator can reduce the cost. In order to further reduce the effect of temperature drift on the amplifier, we put two homemade high-electron-mobility transistors (HEMT) with high saturation power on the 50 K plate in the dilution refrigerator. Additionally, we apply the sixth-order infinite impulse response (IIR) filter in hardware to predistort the Z control signals for each qubit in real-time. This proves useful for the experiment which needs to be repeated continuously over a period of time.

\subsection{Muon detector}

\begin{figure}[ht]
\centering
\includegraphics[width=0.45 \textwidth]{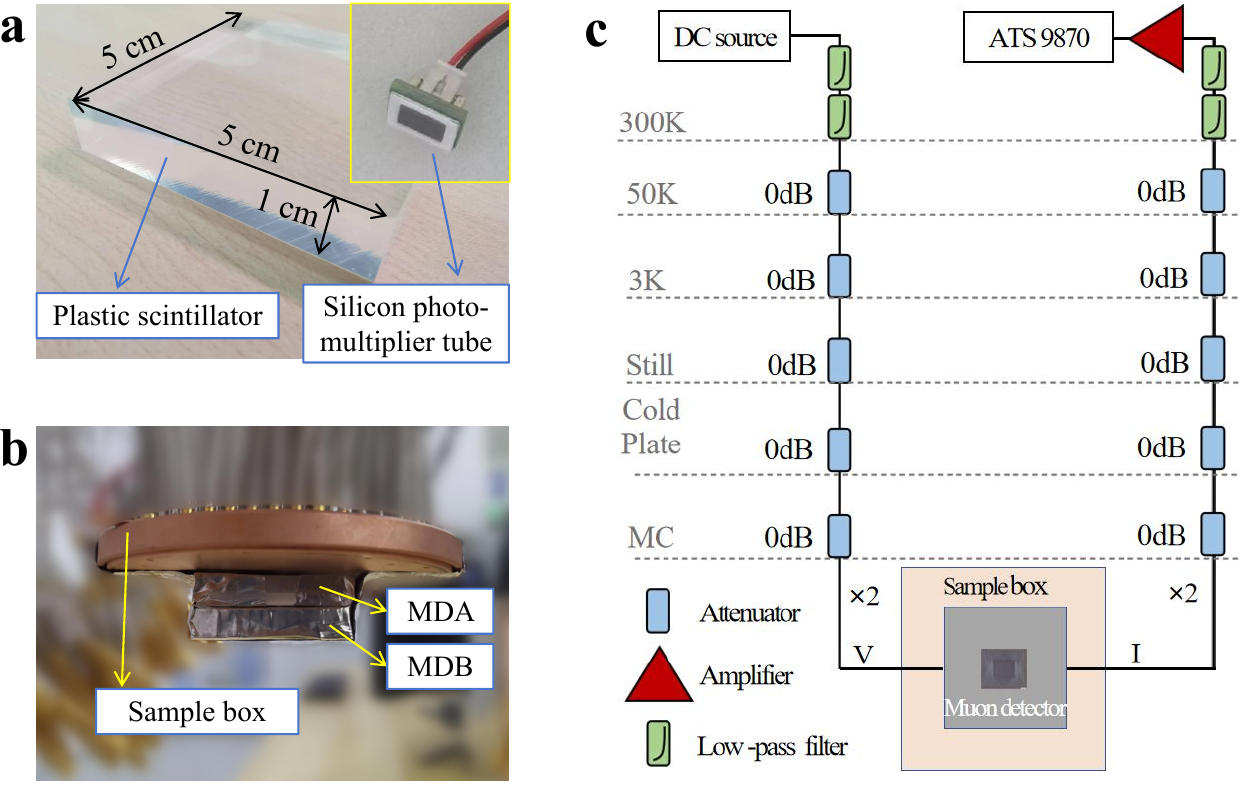}
\caption{
{\bf Muon detector and its operation.} $\textbf{a}$ The muon detector has two main parts: a plastic scintillator and silicon photo-multiplier tube (SiPM). $\textbf{b}$ Two homemade muon detectors, named MDA and MDB, are installed directly beneath the sample box. $\textbf{c}$ The measurement circuits for muon detection.}
\label{fig:muon detector}
\end{figure}

Figure~\ref{fig:muon detector}a illustrates the homemade muon detector with two main components: the plastic scintillator and the silicon photo-multiplier (SiPM) tube. The plastic scintillator will emit light when irradiated. These emitted photons can be captured by the SiPM and amplified to produce an electrical signal.
The SiPMs are affixed to the side of the plastic scintillator. To guarantee that sufficient photons are emitted into the SiPM while preventing external photons from entering, we wrap the scintillator in reflective tin foil and apply optical gel to connect the plastic scintillator and the SiPM. We use two such muon detectors, named MDA and MDB, directly positioned beneath the sample box, as illustrated in Figure~\ref{fig:muon detector}b. Here, the utilization of two muon detectors is to distinguish muons from other high-energy particles and we identify muon events by observing the coincidence between the scintillation signals of both muon detectors.

In Figure~\ref{fig:muon detector}c,  we show the measurement setup of the muon detection. A DC power supply is employed to offer the required voltage to the SiPM. The output signal from the SiPM is further amplified by a room-temperature amplifier and then captured by a data acquisition card (ATS9870). We use two low-pass filters to effectively reduce all kinds of high-frequency noise across the entire dilution refrigerator system, thereby enhancing the efficiency of muon detection. To ensure no muon events are missed, we maintain continuous signal capture at a sampling rate of 20 MHz for the MQSCPJ experiment and 50 MHz for the MQSBF experiment. However, due to the huge amount of data generated in the 6-hour MQSBF experiment (about 2 TB), we only record the signal peaks exceeding a certain threshold. All original data is recorded in the MQSCPJ experiment, resulting in a dataset of $\sim$100 GB.

\subsection{Lead shield and $\gamma$-ray detector}

\begin{figure}[t]
	\centering
	\includegraphics[width=0.45 \textwidth]{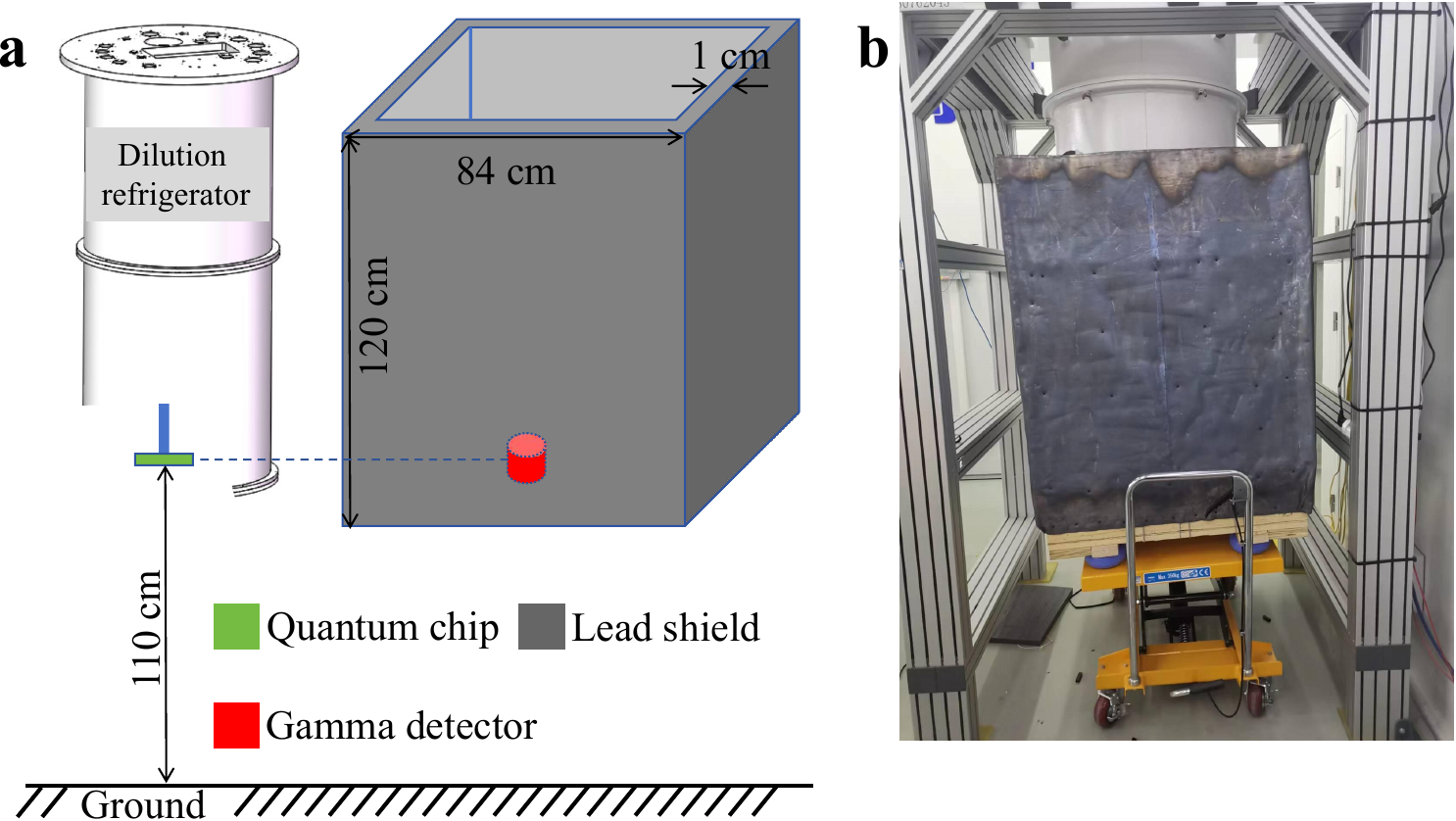}
	\caption{{\bf Lead shield setup.} $\textbf{a}$ Schematic showing the relative locations of the superconducting device, the $\gamma$-ray detector, and the lead shield.
		$\textbf{b}$ Photograph showing the lead shield placed around the outer vacuum chamber (OVC) of the dilution refrigerator.}
	\label{fig:pb shield}
\end{figure}

We have designed a lead shield which can be placed around the outer vacuum chamber (OVC) of the dilution refrigerator, in order to verify that non-muon-induced QP bursts are primarily caused by $\gamma$-rays. Figure~\ref{fig:pb shield}a shows the schematic of the dimension and the relative position of the lead shield, and a photograph with the shield on is shown in Figure~\ref{fig:pb shield}b. The lead shield has a side-door for easier handling in the experiment. To assess the effectiveness of the lead shield, we measure the average time of $\gamma$-ray occurrence with and without the lead shield at the same relative location as the superconducting device. This is accomplished using a commercial $\gamma$-ray detector at room temperature.

In the present experiment, the $\gamma$-ray detector (type CEM DT-9501) is purchased from the JINGDONG online shop (https://item.m.jd.com/product/100019254098.html). The detector can measure various types of radiation, including $\alpha$-rays, $\beta$-rays, X-rays, and $\gamma$-rays. The radiation dose rate ranges from 0.01 $\mu$Sv/hr to 1000 $\mu$Sv/hr, with a pulse dose rate range of 0-4000 cpm or 0-4000 cps. The accumulated radiation dose value can be measured from 0.001 $\mu$Sv to 9999 Sv, and the accumulated pulse dose value ranges from 0 to 9999. The accuracy of the detector is $\pm$10 $\%$ for radiation levels below 500 $\mu$Sv/hr and $\pm$20$\%$ for levels above 600 $\mu$Sv/hr. The sensitivity is 108 pulses or 1000 cpm per mR/hr under Cobalt-60 radiation with a dose rate of 1 $\mu$Sv/hr. 

\section{Experimental methods}

\subsection{QP tunneling}

\begin{figure}[t]
	\centering
	\includegraphics[width=0.4 \textwidth]{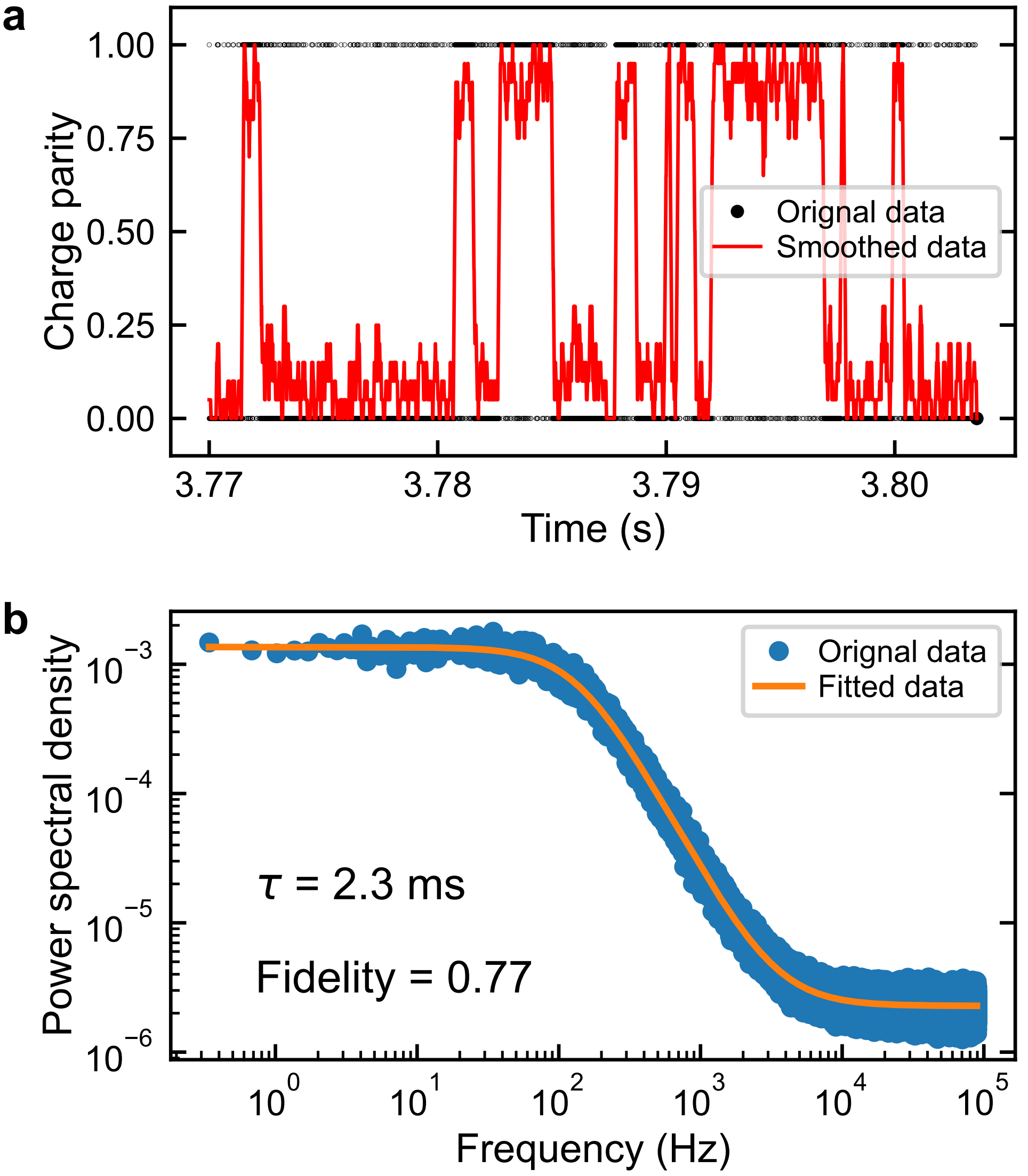}
	\caption{
		{\bf QP tunneling measurement.}  $\textbf{a}$ Time slices of the raw data (black dots) and smoothed data (red line) of charge parity. $\textbf{b}$ The power spectral density (PSD) obtained by Fourier transforming the charge-parity data collected continuously over several seconds, which is fitted by the Lorentzian function (orange line) to give the characteristic time $\tau$ and detection fidelity. Source data are provided as a Source Data file.}
	\label{fig:RTS spectrum}
\end{figure}

As discussed in the main text, the qubit frequencies are significantly different between two-qubit charge-parity states (even or odd) at the charge-sensitive point. By employing a Ramsey-based sequence as shown in Fig.~1b in the main text, we can map the even or odd charge-parity state to the qubit state ($\ket{0}$ or $\ket{1}$). Figure~\ref{fig:RTS spectrum}a shows a typical time slice of continuously monitored charge-parity states. The raw data are represented by black dots, and the red line is the data smoothed by convolution with a square window of 20 sampling points. Many jumps can be identified, indicating occurrences of QP tunnelings. The time series of random QP tunnelings is a kind of random telegraph signal (RTS) ~\cite{Rist__2013_Supp}. By performing a fast Fourier transform on the raw data, we can get the power spectral density (PSD) shown in Figure~\ref{fig:RTS spectrum}b, which can be fitted by the equation  
\begin{equation}
s(f) = \frac{4F^2}{4/\tau+(2\pi f)^2\tau}+(1-F^2)\Delta t ,
\label{equ:qp recombiantion}
\end{equation}
where the variable $f$ represents the frequency of the power spectral density, and $F$ corresponds to the detection fidelity of charge-parity. The parameter $\tau$ represents the average occurrence time of QP tunnelings, and $\Delta t$ denotes the time interval between neighboring data points. The experimental data are depicted as blue dots and the fitted data are depicted as the orange line, which gives $\tau \sim 2.3$ ms, with a detection fidelity of $\sim$~0.77. In this work, we are unable to control the offset charge of individual qubits, resulting in random drift of the offset charge within our qubit system. Hence, we only choose qubits with suitable offset charges and thus high-performance RTS spectrum, determined by the detection fidelity exceeding 0.6 and tunneling time exceeding 1 ms for the MQSCPJ experiment.

\subsection{Thermal effect from QP injection}

\begin{figure}[ht]
\centering
\includegraphics[width=0.5 \textwidth]{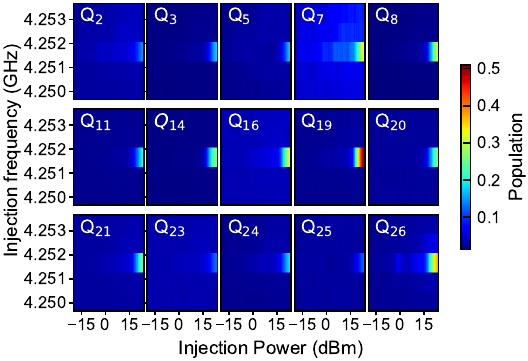}
\caption{
{\bf Determination of QP injection frequency.} For each qubit, the population is measured while varying the injection frequency and power on the readout resonator of the injection qubit. Excitation of each qubit can be observed when the injection frequency matches the bare frequency of the readout resonator of the injection qubit. Source data are provided as a Source Data file.}
\label{fig:qp injection frequency and power}
\end{figure}

In the QP injection experiment, the strong injected microwave pulse at the bare cavity frequency of the injection qubit may have additional effects, such as heating of the qubit chip. To evaluate the impact of this effect on our experiment, we monitor the thermal excitation of each qubit while varying the injection frequency and power of the injection microwave pulse. As shown in Figure~\ref{fig:qp injection frequency and power}, the excitation of each qubit is observed when the frequency of the injected microwave pulse matches the bare readout resonator frequency (4.25 GHz) along with the injection microwave power reaching a sufficiently strong power. First, each qubit is electrically isolated from the injection qubit to prevent direct propagation of generated QPs to all measurement qubits. Second, minimal excitation is observed when biasing the injection microwave frequency away from the bare readout resonator frequency, even at the highest microwave injection power. This indicates that the slight heating added to the qubit chip does not excite the qubit. Finally, nearly identical excitation behavior and similar transition points in the average charge-parity time $\tau$ and energy relaxation time $T_1$ for each qubit (as shown in Fig.~2a and Fig.~2b in the main text) strongly indicate that the excitation of each qubit is predominantly influenced by the propagation of numerous phonons in the substrate resulting from the recombination of injected QPs.  

\subsection{QP recombination rate and QP trapping}

\begin{figure}[t]
	\centering
	\includegraphics[width=0.45 \textwidth]{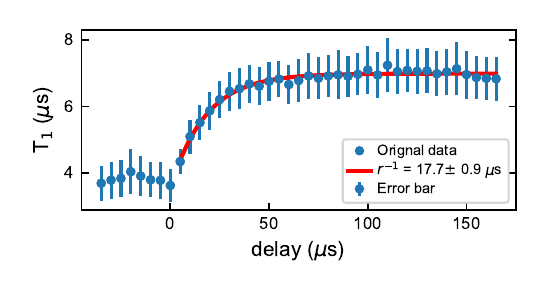}
	\caption{
		{\bf QP recombination time.}  The energy relaxation time $T_1$ is measured by varying the time delay after the injection of QPs. The data are fitted to give the QP recombination time 1/$r$. The red solid line is the fit. The error bars represent the standard deviations of five replicate experiments. Source data are provided as a Source Data file.}
	\label{fig:qp injection delay}
\end{figure}

We measure the recombination rate of the QPs near the qubit junctions. The dynamics of QP density, denoted as $\rho_\text{qp}$, is described by the following equation~\cite{Wang_2014_Supp}:
\begin{equation}
d\rho_\text{qp}/dt = -r\rho_\text{qp}^2-s\rho_\text{qp}+\rho_\text{qp0} ,
\label{equ:qp recombiantion}
\end{equation}
where $r$ represents the rate of recombination into Cooper pairs, $s$ is the rate of trapping or escaping of QPs near the qubit junctions, and $\rho_\text{qp0}$ denotes the background density of QPs generated thermally or by stray photons. On the other hand, the energy relaxation rate of the qubit is 
\begin{equation}
\gamma_{01} = \frac{\rho_\text{qp}}{\pi}\sqrt{2\Delta\omega_{01}},
\label{equ:qp T1}
\end{equation}
where $\Delta$ is the energy gap of the superconducting films of the qubit junction, and $\omega_{01}$ is the qubit frequency. Therefore, we can obtain the recombination rate of QPs by measuring the energy relaxation rate $\gamma_{01}$. To this end, we inject the excess QPs lasting for a controlled time followed by a delay time after the injection. Then, we measure the energy relaxation rate $\gamma_{01}$ of the qubit. The energy relaxation time $T_{1} = 1/\gamma_{01}$ as a function of the time delay is shown in Figure~\ref{fig:qp injection delay}. The result (blue dots) is fitted by the exponential function shown as a red solid line. The fitted recombination time is about 17.7 $\mu$s, nearly the same as the relaxation time of QP burst detected by MQSBF in Fig.~3h of the main text. 

The observed QP recombination time is two to three orders of magnitude shorter than that in Google's experiment ($\sim$~25 milliseconds)~\cite{mcewen_resolving_2022}. This can be explained by higher density of QPs trapped in the Al films of the qubit junctions. As can be seen in Fig.~1a and Fig.~5 in the main text, the qubit capacitor pads are made of Ta films and the QP traps form due to the superconducting gap difference between Al (0.18 meV) and Ta (0.69 meV), similar to the situation discussed in Ref.~\cite{fink2024superconducting}. The trapped high-density QPs will lead to high tunneling events~\cite{Booth1987} leading to high charge-parity jump and bit-flip events in our experiment. To estimate the QP density $\rho_\text{qp}$, we consider the QP recombination rate $r$ given by~\cite{martinis2021saving, mcewen_resolving_2022}:
\begin{equation}
r \approx x_\text{qp} \cdot21.8/\tau_{0},
\label{equ:qp recombianation rate}
\end{equation}
where $x_\text{qp} = \rho_\text{qp}/\rho_\text{cp}$ is the normalized density of QPs. The density of Cooper pairs $\rho_\text{cp}$ in Al film is $\sim4.18\times10^{6} \mu m^{-3}$, and $\tau_0$~=~440 ns is the characteristic QP relaxation time for Al. From the equation, high QP density will lead to fast QP recombination rate. In the present experiment, we estimate $x_\text{qp}\sim5.6\times10^{-4}$ considering the recombination time of $\sim 36~\mu$s from the MQSCPJ experiment, which leads to a $\rho_\text{qp} \sim$ 2344 $\mu$m$^{-3}$. If we take the volume of an Al film trap to be $\sim$ 3 $\mu$m$^3$, the number of QPs in the trap will be around 7032. Corresponding values in Google's experiment are $x_\text{qp}\sim8.07\times10^{-7}$ and $\rho_\text{qp} \sim$ 3.37 $\mu$m$^{-3}$. If we assume the background power levels, such as the pair-breaking power from far-infrared thermal background, are comparable in the two experiments, the data would give rise to a factor of $\sim$~695 increase of QP density resulting from QP trapping.

\subsection{Detection of QP bursts}

In order to detect QP bursts, we employ two methods in our experiment. The first is to continuously monitor multiqubit simultaneous charge-parity jumps (MQSCPJ) and the second is to continuously monitor multiqubit simultaneous bit flips (MQSBF). When QP burst occurs in the qubit array, it simultaneously raises the rate of QP tunneling across multiple qubits, which leads to simultaneous changes in the charge-parity state when an odd number of QPs tunnel, or influences the energy relaxation of multiple qubits if the frequency difference between the initial and final QP states matches the frequency of the qubit.

\subsubsection{Correlated multiqubit charge-parity jumps}

\begin{figure*}[t]
\centering
\includegraphics[width=0.9 \textwidth]{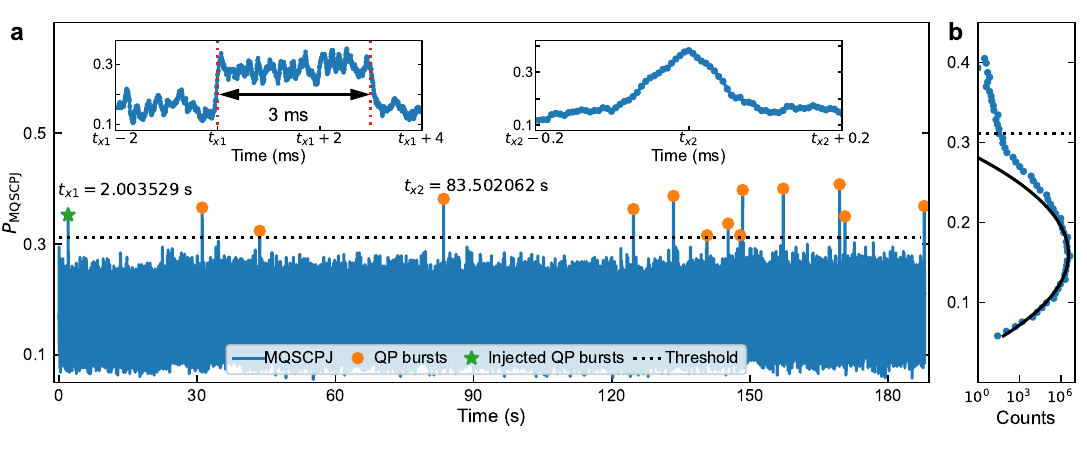}
\caption{{\bf QP bursts detected by MQSCPJ method.} $\textbf{a}$ The smoothed MQSCPJ ratio $P_\mathrm{MQSCPJ}$ is continuously monitored over a duration of 187.9 s. The green star and orange dots are identified as QP bursts as they exceed the threshold ($\sim$ 0.31) shown by the black dashed line. The insets provide a zoomed-in view of the QP-burst events. The left one corresponds to the event induced by the injection QPs, and the other is identified as the QP bursts from high-energy events. $\textbf{b}$ The histogram of the $P_\mathrm{MQSCPJ}$.  We fit the data below the threshold via a Gaussian function, as shown by the black line. Source data are provided as a Source Data file.}
\label{fig:MQSCPJ}
\end{figure*}

\begin{figure*}[t]
	\centering
	\includegraphics[width=0.9 \textwidth]{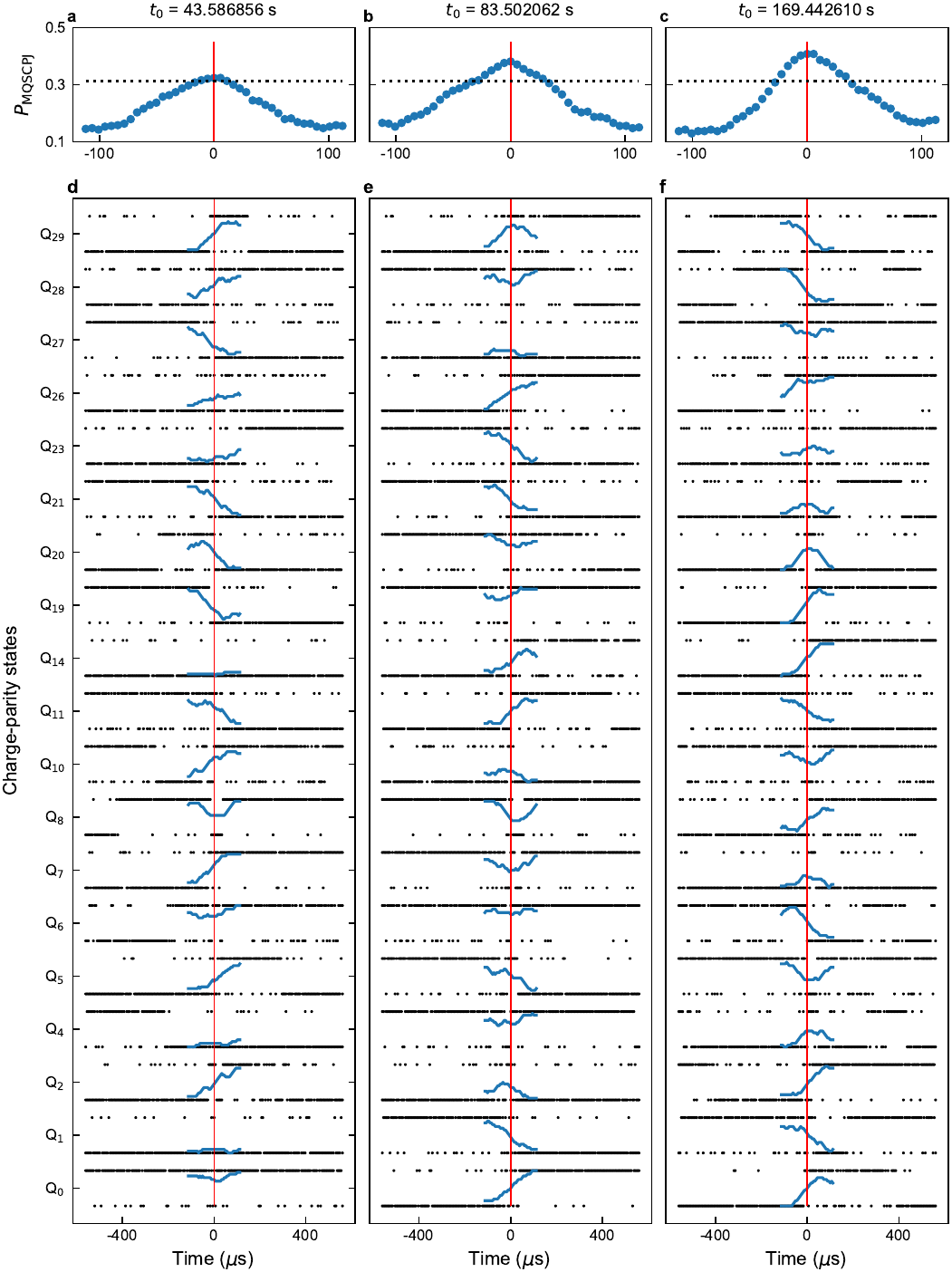}
	\caption{
		{\bf Details in the time domain of three typical QP bursts detected by the MQSCPJ method.} $\textbf{a}$, $\textbf{b}$, $\textbf{c}$ The zoomed-in views of the QP bursts at three times in Figure~\ref{fig:MQSCPJ}a of $\sim$44 s, $\sim$84 s, and $\sim$169 s, respectively. $\textbf{d}$, $\textbf{e}$, $\textbf{f}$ Corresponding time series of charge-parity states for all selected qubits. Their smoothed data within 200 $\mu$s are shown by the blue lines. The red solid lines indicate the times identified as QP bursts. Source data are provided as a Source Data file.}
	\label{fig:zoom-in of MQSCPJ}
\end{figure*}

In our experiment, the initial charge-parity state of each qubit is random. After the mapping process shown in Figs.~1b and c of the main text, this randomness manifests as the qubit in either $\ket{0}$ or $\ket{1}$ state. Since our interest is only in the jump of the charge-parity state (from $\ket{1}$ to $\ket{0}$, or vice versa), the initial randomness does not affect the experimental results. The charge-parity jump rate is determined by the QP tunneling rate, which is generally proportional to the QP density near the junction barrier produced by the QP burst. As for noise sources, there are two primary contributions. One is random noise due to qubit readout or operation errors, and the other is charge-parity jump induced by background QP tunneling. In our experiment, the rate of the background charge-parity jumps is typically below 1 kHz, which is much lower than the charge-parity jump rate caused by high-energy events, as further detailed below. Therefore, the dominant noise is qubit readout or operation errors, which are seen in the raw data as occasional points deviating from the charge-parity states. Based on the signal and noise characteristics of the charge-parity states described above, we apply a square-window convolution to the raw data to improve the signal-to-noise ratio. The chosen window length is 20 sampling point, effectively suppressing random noise with minimal impact on the charge-parity jump ``points''. However, it does broaden the transition edges. While this smoothing may also suppress some “high-frequency” signals originating from high-energy events with higher QP density, such signals will still be relaxed, producing a measurable tail in the data.

In addition, QP bursts generate QPs across multiple qubits simultaneously, increasing their tunneling probabilities in a short time. By averaging this smoothed charge-parity data over multiple qubits, it is possible to further reduce random noise and retain multiqubit correlation signals as much as possible, thereby improving the signal-to-noise ratio to get the $P_\text{MQSCPJ}(n)$ as follows,

\begin{equation}
P_\text{MQSCPJ}(n) = \frac{1}{M} \sum_{m=1}^{M} \left( 0.5 - \left| \sum_{\tau=1}^{L} f_L(\tau) d_m(n - \tau) - 0.5 \right| \right),
\label{equ:MQSCPJ rate}
\end{equation}
where
\begin{equation}
f_L(\tau) = \begin{cases} 
      0 & \text{if } \tau < 1 \\
      1/L & \text{if } 1\leq \tau \leq L\\
      0 & \text{if } \tau > L
   \end{cases} ,
\label{equ:MQSCPJ window}
\end{equation}
here $n$ is the sampling points, $M$ is the total number of selected qubits, $L$ is the length of the square window for smoothing the raw data, $d_m(n)$ is the sequence of $\ket{0}$'s or $\ket{1}$'s corresponding to the charge-parity states of each selected qubit, and $f_L(\tau)$ is the squared window function for smoothing. In Equation~(\ref{equ:MQSCPJ rate}), for a single qubit, the condition for identifying a charge-parity jump is that the quantity $0.5- |\sum_{\tau=1}^{L} f_L(\tau)d_m(n-\tau)-0.5|$ equals 0.5. When this value approaches 0, no charge-parity jumps occur. This quantity is averaged across multiqubits to obtain $P_\text{MQSCPJ}(n)$.

In Figure~\ref{fig:MQSCPJ}a, $P_\text{MQSCPJ}(n)$ is monitored over 187.9 s. Notably, anomalous peaks (orange dots) exceeding the predetermined threshold of 0.31 (black dashed line) are observed. The threshold is set to the mean of all data plus 6 times the standard deviation. This setting leads to the average number of peaks caused by random fluctuations exceeding the threshold being kept around 0.1 during the monitoring time of $\sim$ 599 s. However, 13 anomalous peaks are observed in this experiment. The corresponding histogram of $P_\text{MQSCPJ}(n)$ is presented in Figure~\ref{fig:MQSCPJ}b. We can observe a significant departure from the Gaussian distribution. These findings indicate the presence of QP bursts in our quantum processor. The detailed shapes of the anomalous peaks are depicted in the insets. The left inset correspond to the events induced by the injected QPs, lasting for a controlled 3 ms, while the right inset are the anomalous events induced by the high-energy particles.

\begin{figure*}[!tp]
	\centering
	\includegraphics[width=0.75 \textwidth]{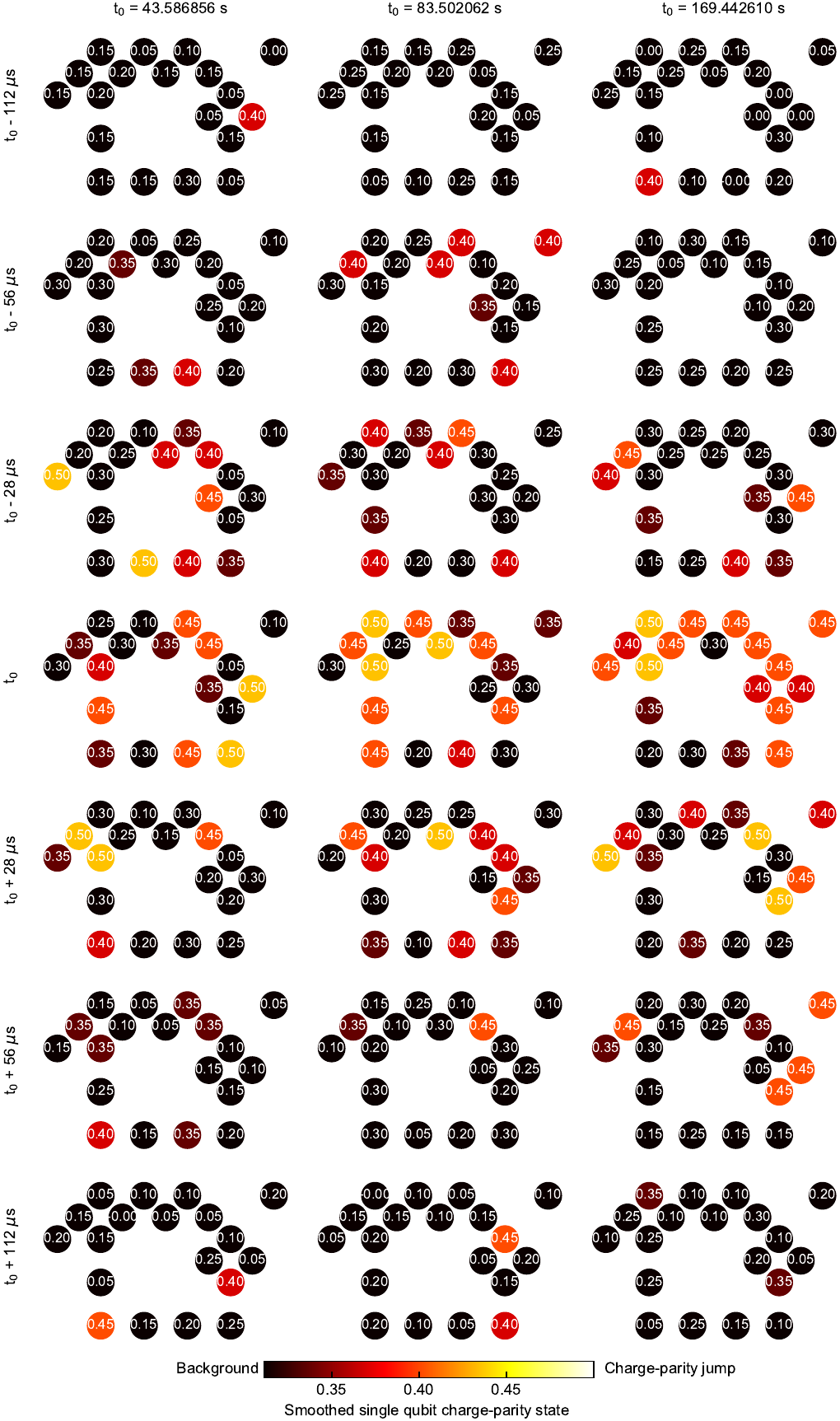}
	\caption{{\bf Details in the space domain of three typical QP bursts detected by the MQSCPJ method.} For each QP burst, 7 distributions on multiple qubits in the space domain are shown, corresponding to 7 different times before and after the times shown in Figure~\ref{fig:zoom-in of MQSCPJ}. Source data are provided as a Source Data file.}
	\label{fig:zoom-in of MQSCPJ in space}
\end{figure*}

The three typical QP bursts in Figure~\ref{fig:MQSCPJ}a are magnified and shown in Figures~\ref{fig:zoom-in of MQSCPJ}a, \ref{fig:zoom-in of MQSCPJ}b, and \ref{fig:zoom-in of MQSCPJ}c, respectively. Figures~\ref{fig:zoom-in of MQSCPJ}d, \ref{fig:zoom-in of MQSCPJ}e, and \ref{fig:zoom-in of MQSCPJ}f separately show the raw data (black dots) and the corresponding smoothed data (blue lines) calculated by $\sum_{\tau=1}^{L} f_L(\tau)d_m(n-\tau)$ for each selected qubit in these three QP bursts. The solid red lines indicate the moment of a QP burst. Here we can see that most of the selected qubits undergo either charge-parity jumps or frequent transitions between different charge-parity states. 

\begin{figure*}[t]
	\centering
	\includegraphics[width=0.7 \textwidth]{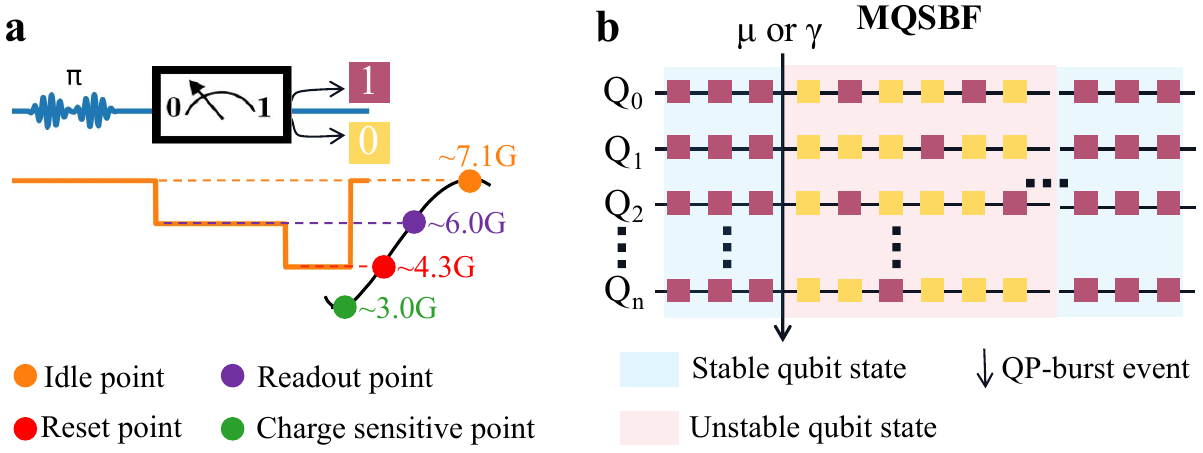}
	\caption{{\bf Measurement of multiqubit simultaneous bit flip.} $\textbf{a}$ The measurement sequence to determine the qubit bit flip from $\ket{1}$ state to $\ket{0}$ state. We bias the qubit to the idle point, readout point, and reset point of the qubit spectrum for qubit state preparation, measurement, and initialization, respectively. $\textbf{b}$ Schematic of one typical QP burst. We continuously monitor the bit flip across multiple qubits within a period of 5.6 $\mu$s and a QP burst is identified when we observe a multiqubit
		simultaneous bit fip (MQSBF).}
	\label{fig:sequence of MQSBF}
\end{figure*}

\begin{figure*}[b]
	\centering
	\includegraphics[width=0.9 \textwidth]{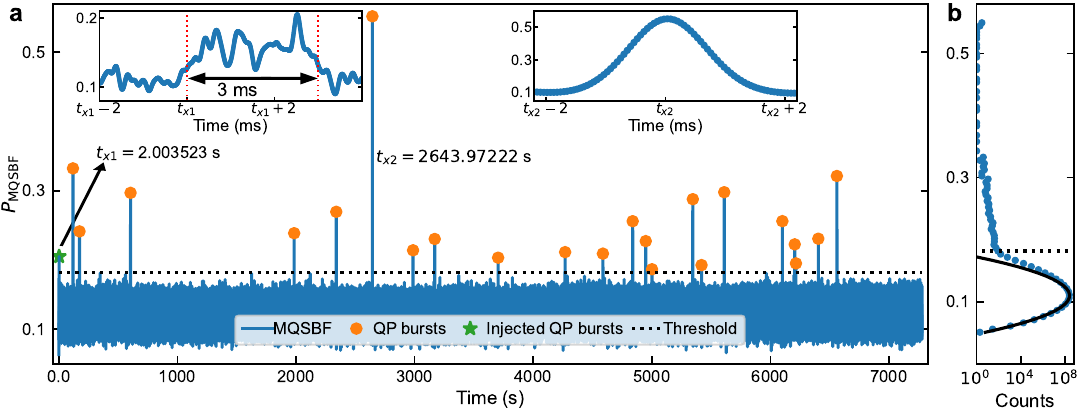}
	\caption{
		{\bf QP bursts detected by the MQSBF method.} $\textbf{a}$ The smoothed MQSBF ratio $P_\mathrm{MQSBF}$ is continuously monitored over a duration of 7279.9 s. The selected peaks (green star and orange dots) are identified as the QP bursts, and the threshold is shown as a black dashed line. The insets provide the zoomed-in views of two QP bursts. The left one corresponds to the event induced by the injected QPs, and the other is from a high-energy event. $\textbf{b}$ The histogram of $P_\mathrm{MQSBF}$ with the Gaussian fit on data below the threshold, as shown by the black dashed line. Source data are provided as a Source Data file.}
	\label{fig:MQSBF}
\end{figure*}

The spatial distribution of selected qubits in these three QP bursts, as determined by the quantity of $0.5- |\sum_{\tau=1}^{L} f_L(\tau)d_m(n-\tau)-0.5|$, provides insights into the behavior of QP burst for each qubit. In Figure~\ref{fig:zoom-in of MQSCPJ in space}, we display 7 time-slices before and after the occurrence of these three QP bursts. We observe that the majority of qubits become illuminated and then return to their original states within about 100 $\mu$s. Additionally, as the peak value of P$_\text{MQSCPJ}(n)$ increases, more qubits become illuminated.

\subsubsection{Correlated multiqubit bit flips}

This method involves measuring the simultaneous energy relaxation induced by QP bursts across multiple qubits. The measurement sequence for a single qubit is shown in Figure~\ref{fig:sequence of MQSBF}a. The qubit is initially prepared in the excited state $\ket{1}$, followed by a short idle time of 300 ns prior to the qubit measurement. Since the average energy relaxation time of the qubit in Table~\ref{tab:all parameters} is 9.1 $\mu$s, the qubit has a high probability close to 1 in the $\ket{1}$ state. However, during a QP burst, this probability significantly decreases, leading to the state $\ket{1}$ easily flipping to state $\ket{0}$, which will result in a bit-flip error. The dominant noise in the bit-flip experiment comes from the readout error of the qubit, which is manifested as the occasional $\ket{0}$ in the excited state $\ket{1}$ and the occasional $\ket{1}$ in the ground state $\ket{0}$. Considering the above characteristics of the signal and noise in the bit-flip experiment, we use a Gaussian window to smooth the raw data instead of using a square window in the charge-parity experiment. In most cases, this single-qubit data analysis has difficulty extracting signals under background noise. Therefore, we also average the smoothed data from multiple qubits to suppress random noise, as shown in Figure~\ref{fig:sequence of MQSBF}b. 

To quantify the MQSBF experiment, we quantize the above process by defining the smoothed MQSBF ratio, which is the ratio of the number of error qubits to the total number of qubits. Mathematically, the smoothed MQSBF ratio, $P_\text{MQSBF}(n)$, is defined as
\begin{equation}
P_\text{MQSBF}(n) = \sum_{\tau=-4L}^{4L} f_L(\tau)\lbrace\frac{1}{M}\sum_{m=1}^{M}[1- d_m(n-\tau)])\rbrace ,
\label{equ:MQSBF rate}
\end{equation}
where
\begin{equation}
f_L(\tau) = \begin{cases} 
      0 & \text{if } \tau < -4L \\
    \frac{1}{\sqrt{2\pi}L}\exp(-\frac{\tau^2}{L^2})& \text{if } -4L\leq \tau \leq 4L\\
      0 & \text{if } \tau > 4L
   \end{cases}.
\label{equ:MQSBF window}
\end{equation}

In Figure~\ref{fig:MQSBF}a, 22 anomalous peaks (orange dots) exceeding the established threshold of 0.18 (black dashed line) have been observed. The threshold is set to the mean of all data plus 6.6 times the standard deviation. This can ensure that the average number of peaks caused by random fluctuations exceeding the threshold remains $\sim$ 0.1, during the monitoring time of $\sim$ 6 hr in the MQSBF experiment.

\begin{figure*}[t]
	\centering
	\includegraphics[width=0.9 \textwidth]{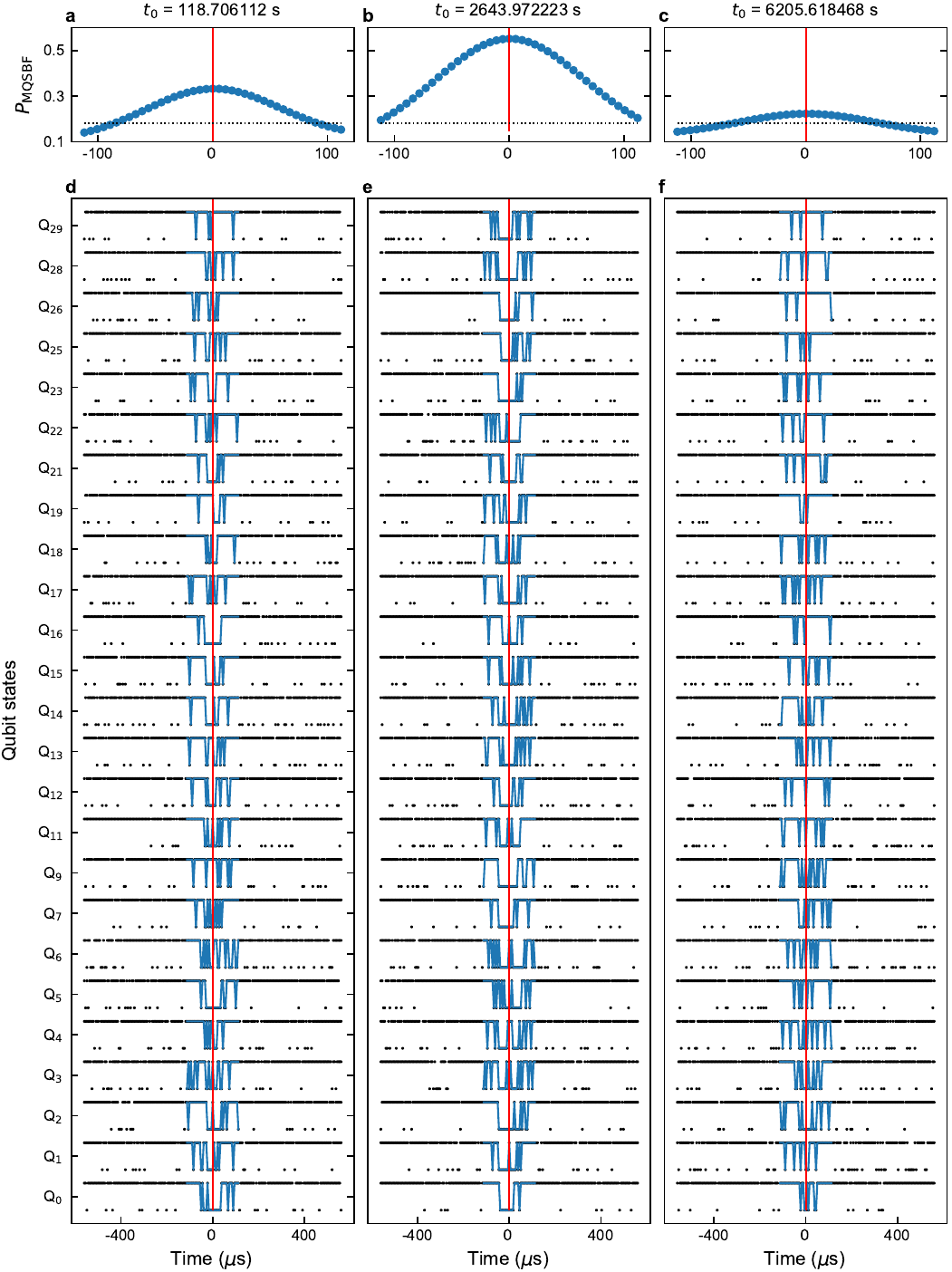}
	\caption{
		{\bf Details in the time domain of three typical QP bursts detected by MQSBF method.} $\textbf{a}$, $\textbf{b}$, $\textbf{c}$ The zoomed-in views of smoothed MQSBF ratio $P_\mathrm{MQSBF}$ of QP bursts as shown in Figure~\ref{fig:MQSBF}a at three different times. $\textbf{d}$, $\textbf{e}$, $\textbf{f}$ Time series of the states of the selected qubits. The blue lines highlight the multiqubit simultaneously flip from $\ket{1}$ state to $\ket{0}$ state with high probability. The red solid lines indicate the times when QP burst occurs. Source data are provided as a Source Data file.}
	\label{fig:zoom-in of MQSBF}
\end{figure*}

\begin{figure*}[!tp]
	\centering
	\includegraphics[width=0.78 \textwidth]{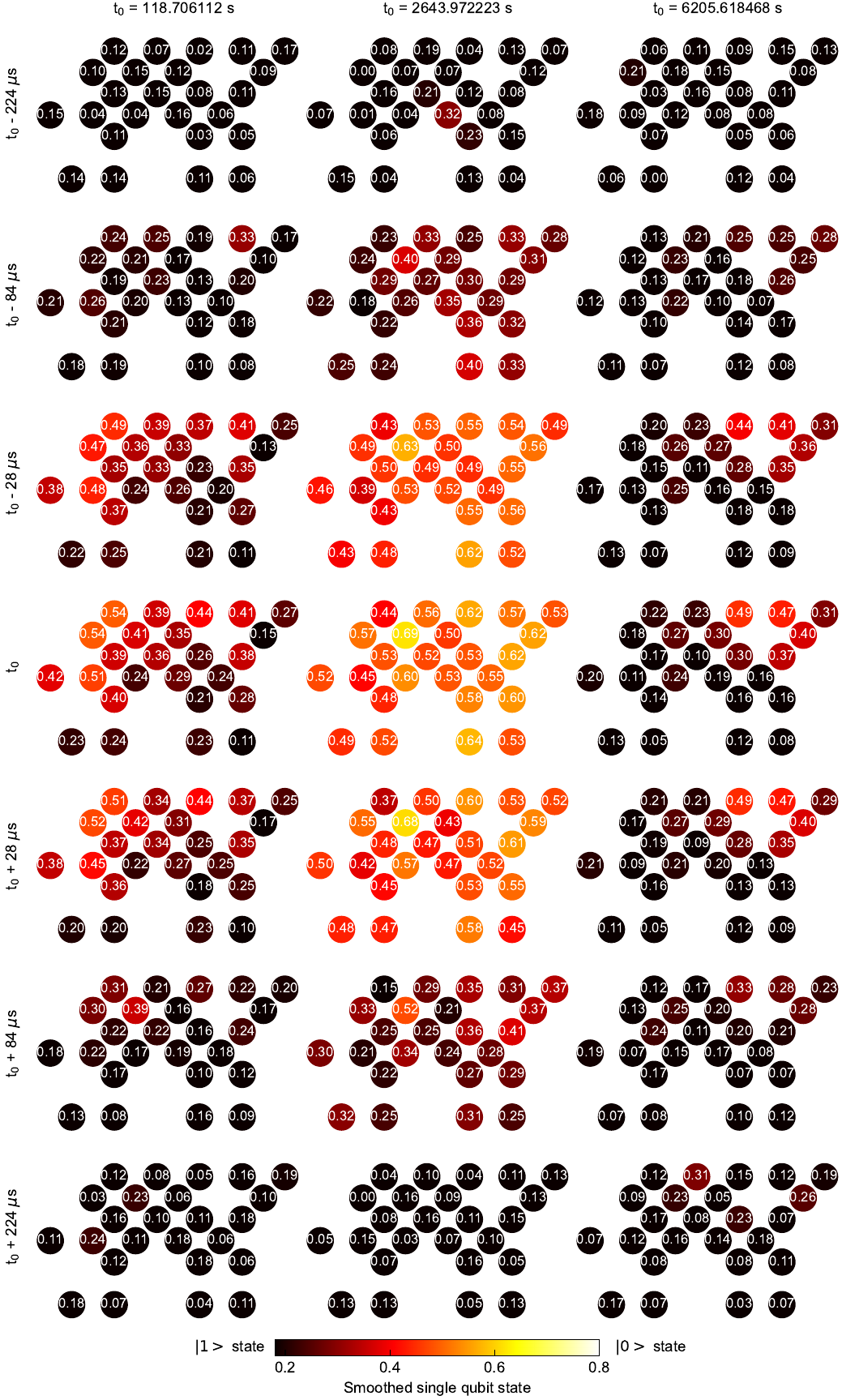}
	\caption{
		{\bf Details in the space domain of three typical QP bursts detected by the MQSBF method.} For each QP burst, 7 distributions on multiple qubits in the space domain are shown, corresponding to 7 different times before and after the times shown in Figure~\ref{fig:zoom-in of MQSBF}. Source data are provided as a Source Data file.}
	\label{fig:zoom-in of MQSBF in space}
\end{figure*}

Similar to the MQSCPJ experiment, we also show the time distribution details of three typical QP bursts in Figure~\ref{fig:zoom-in of MQSBF}a, Figure~\ref{fig:zoom-in of MQSBF}b and, Figure~\ref{fig:zoom-in of MQSBF}c. At the same time, the black dots in Figure~\ref{fig:zoom-in of MQSBF}d, Figure~\ref{fig:zoom-in of MQSBF}e, and Figure~\ref{fig:zoom-in of MQSBF}f give the raw data corresponding to these QP-burst events, while the blue lines show the behavior of each qubit during a QP burst with a duration of about 200 $\mu$s, where the states of many selected qubits flip to the $\ket{0}$ with high probability. The solid red lines in these figures indicate the QP-burst times selected according to the MQSBF method. We also show the detail spatial distribution of the selected qubits of the three QP-burst events at 7 times before and after the burst times in Figure~\ref{fig:zoom-in of MQSBF in space}. The temporal and spatial distributions of these QP bursts provide some simple insights into how QP bursts occur, propagate, and eventually recover in a qubit array. We also find that, in Figure~\ref{fig:zoom-in of MQSBF}e, for the higher peaks, almost all qubits are more affected by the QP burst.

\subsection{Threshold determination}

\begin{figure*}[t]
\centering
\includegraphics[width=0.9 \textwidth]{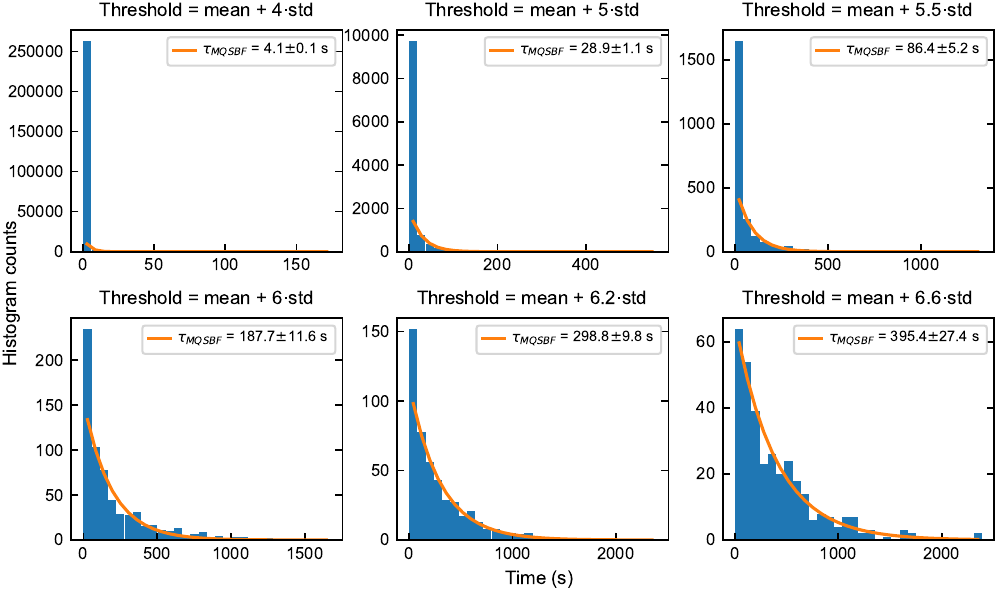}
\caption{
{\bf The extracted average occurrence time of QP bursts detected by MQSBF method using different thresholds.} As we increase the threshold, the average occurrence time $\tau_\text{MQSBF}$ will increase, but the height of the first bar on the left will decrease, indicating that the proportion of random noise exceeding the threshold is decreasing. Source data are provided as a Source Data file.}
\label{fig:threshold}
\end{figure*}

\begin{figure*}[t]
	\centering
	\includegraphics[width=0.55 \textwidth]{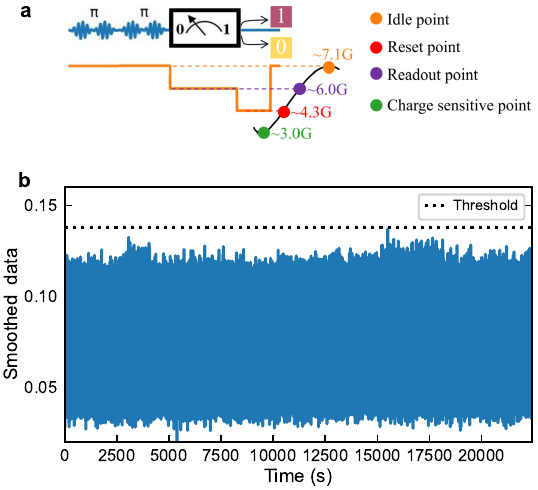}
	\caption{
		{\bf Monitoring qubit operation and readout errors over long periods.} $\textbf{a}$ The sequence for the experiment. Similar to bit-flip experiment shown in Figure~\ref{fig:sequence of MQSBF}, the qubit is initialized to the ground state by using two consecutive $\pi$ pulses, but without waiting for an additional time. We then measure the qubit state and reset the qubit to the ground state. $\textbf{b}$ The measurement sequence is performed continuously on multiple qubits for 22400 s. The acquired data are then analyzed using a method similar to the MQSBF experiment, with a threshold (black dashed line) set to 6.6 sigma ($\sim$ 0.14). After setting this threshold, no abnormal peaks are found that exceed the threshold. Source data are provided as a Source Data file.}
	\label{fig:stability of qubits operation and readout}
\end{figure*}

In the MQSCPJ or MQSBF experiment, the identification of QP bursts is accomplished by comparing the peak value of the smoothed MQSCPJ or MQSBF ratio with a given threshold. The selection of an appropriate threshold is important in ensuring accurate event detection. In this work, we set the thresholds to be the means of all data plus 6 and 6.6 times the standard deviation for the smoothed MQSCPJ and MQSBF ratios, respectively. Nonetheless, we note that an excessively high threshold may result in the overlooking of certain QP bursts, while an exceedingly low threshold may introduce excessive noise events. As illustrated in Figure~\ref{fig:threshold}, the higher the threshold we set, the longer the average occurring time we can obtain. Therefore, an optimal balance in threshold selection is essential to ensure accurate and reliable detection of QP bursts.

The primary objective of our experiment is to detect QP bursts in superconducting qubits triggered by high-energy events. To achieve this safely, we adopted a conservative threshold, ensuring that the average random events exceeding the threshold (6 sigma for MQSCPJ and 6.6 sigma for MQSBF) remains about 0.1 during long time measurements.

For the MQSCPJ method, the high sensitivity allows to detect a muon flux of 0.4 min$^{-1}$cm$^{-2}$, even with a high threshold. This is calculated using the qubit chip area and is comparable to that measured by muon detectors (0.506 min$^{-1}$cm$^{-2}$) when an effective area of the qubit array is considered instead of the qubit chip area (see below). The MQSCPJ method shows almost no detection loss for muon events under the selected threshold.

In contrast, the MQSBF experiment, with lower sensitivity, requires extended measurement time (several hours). To maintain accuracy under some drifts caused by factors like TLS or environmental electrical noise, we use a higher threshold. This may miss some QP bursts.

\section{Extended data}

\subsection{Correlated qubit excitations}

Figure~\ref{fig:qp injection frequency and power} demonstrates that the injected QPs can have a noticeable impact on the qubit excitation. Additionally, the high QP density may also induce variations in the qubit frequencies, thus potentially leading to the single-qubit gate error. We perform the following measurements to investigate the phenomenon in our experiment. Instead of preparing the qubit at the $\ket{1}$ state in the MQSBF experiment, we apply two consecutive $\pi$ pulses to excite the qubits to the state $\ket{1}$ and then back to $\ket{0}$, as shown Figure~\ref{fig:stability of qubits operation and readout}a. We continuously monitor this on all selected qubits for a duration of 22400 s, with a period of 5.6 $\mu$s, as illustrated in Figure~\ref{fig:stability of qubits operation and readout}b. After setting the threshold to 6.6 sigma (similar to the MQSBF experiment), no anomalous peaks exceeding the threshold were found, indicating no correlated qubit excitations. 
The result also confirms the effectiveness and reliability of the MQSCPJ and MQSBF experiments.

\subsection{Details of coincident events}

\begin{figure}[t]
	\centering
	\includegraphics[width=0.47 \textwidth]{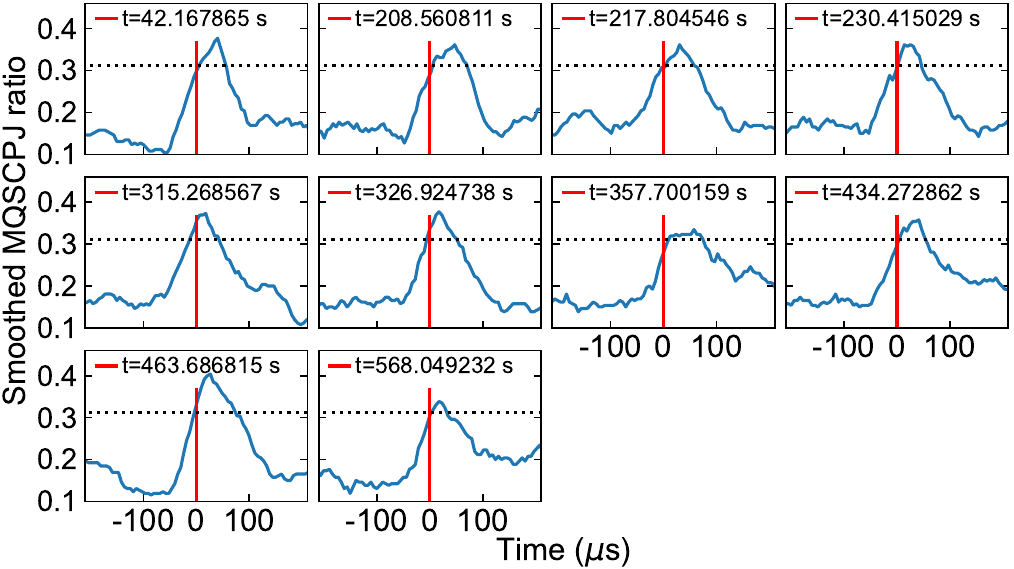}
	\caption{
		{\bf Relative time location between muon events and QP bursts detected by MQSCPJ.} Zoomed-in views of all 10 coincident events in Fig.~3a of the main text. The blue lines represent the detailed shapes of the QP burst after smoothing and the red lines indicate the times of muon events. The black dashed lines are the thresholds. Source data are provided as a Source Data file.}
	\label{fig:coincidence in MQSCPJ}
\end{figure}

\begin{figure}[!tp]
	\centering
	\includegraphics[width=0.47 \textwidth]{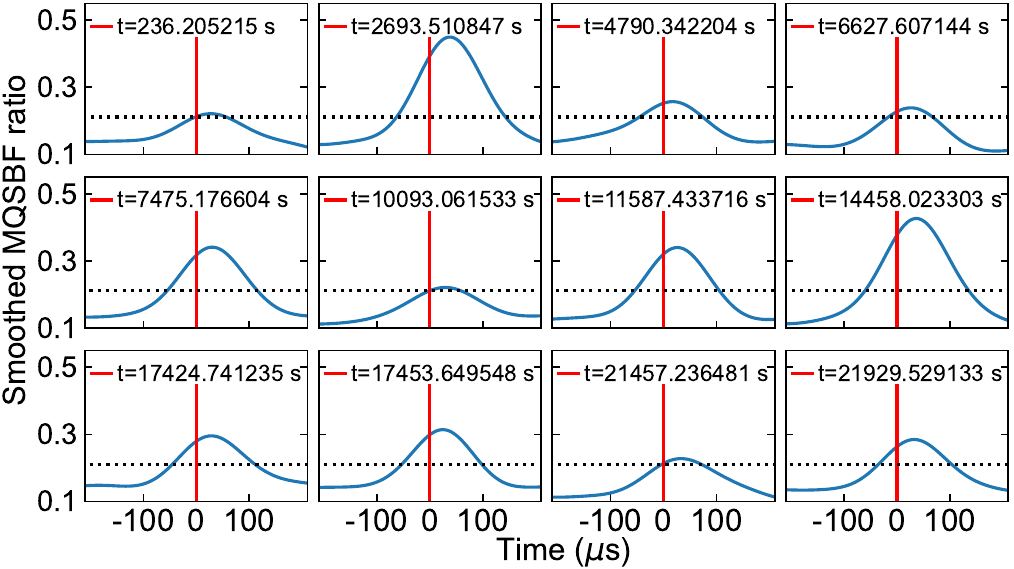}
	\caption{
		{\bf Relative time location between muon events and QP bursts detected by MQSBF.} Zoomed-in view of all 12 coincident events in Fig.~3d of the main text. The blue lines represent the detailed shapes of the QP burst after smoothing and the red lines indicate the times of muon events. The black dashed lines are the thresholds. Source data are provided as a Source Data file.}
	\label{fig:coincidence in MQSBF}
\end{figure}

In Fig.~3 of the main text, we have shown the results of coincident events between muon events and QP bursts detected by the MQSCPJ and MQSBF methods. The details of these coincident events, consisting of 10 and 12 instances respectively, are presented in Figure~\ref{fig:coincidence in MQSCPJ} and Figure~\ref{fig:coincidence in MQSBF} by zoomed-in views. The identification of these simultaneous events relies on a set of rules, namely that the peak of the QP burst and the muon event fall within a specified time window of 100 $\mu$s. In fact, choosing a time window from 50 $\mu$s to 1000 $\mu$s does not have noticeable differences on the results, since the QP bursts in our experiment occur about every 12.3 s. This time scale is too long compared to the window length, so the probability of two QP bursts occurring within the window is very low.

In addition, we find that the smoothed peaks of QP bursts consistently align with the red lines for the times of muon events, with slight time delay in Figure~\ref{fig:coincidence in MQSCPJ} and Figure~\ref{fig:coincidence in MQSBF}, thereby confirming they are caused by a single muon penetrating the superconducting qubit device and muon detector. The slight time delay should result from the complex process to convert the deposited energy in the substrate to QP burst. Since the time delay remains almost unchanged in different coincident events, we can use the muon events as triggers to average all these muon-induced QP bursts, as depicted in Fig.~3g and Fig.~3h of the main text.

\section{Measurement of muon flux}

\begin{figure*}[t]
	\centering
	\includegraphics[width=0.9\textwidth]{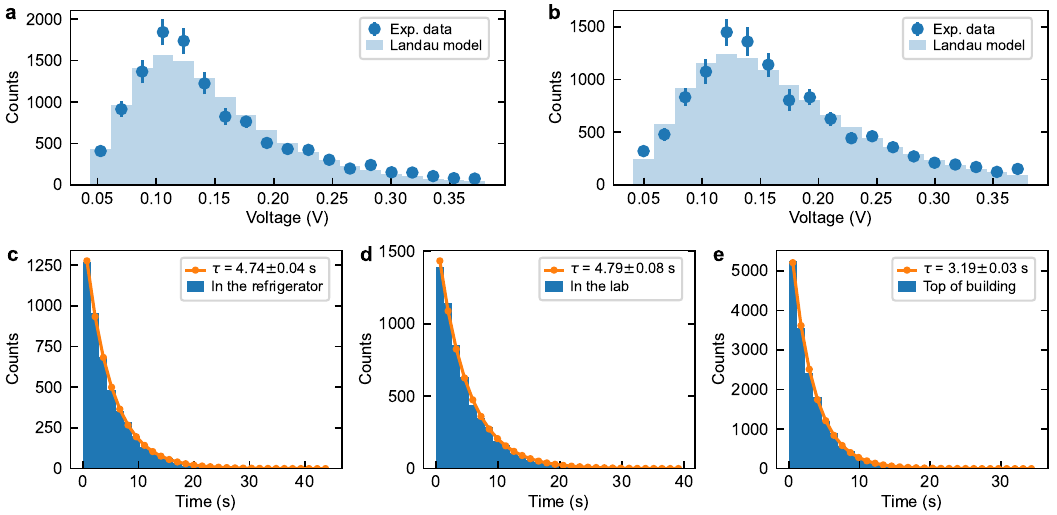}
	\caption{\label{fig:trigger}{\bf Measurement of muon flux under different conditions.} $\textbf{a}$, $\textbf{b}$ The distributions of the voltage signals of MDA and MDB detectors (blue dots) measured in the refrigerator at mK temperature are fitted with Landau distribution (light blue bars).  $\textbf{c}$, $\textbf{d}$, $\textbf{e}$ The histograms of the time intervals between the neighboring muon events measured in the refrigerator, in the laboratory, and on the top of the laboratory building, respectively, which are exponentially fitted to determine the respective average occurrence time $\tau$. Source data are provided as a Source Data file.}
\end{figure*}

In order to measure the muon flux under different conditions, we model the distribution of measured voltage signals of the muon detector with the Landau distribution~\cite{Vavilov1957}. The results for the two detectors placed below the sample box in the refrigerator at mK temperature are shown in Figure~\ref{fig:trigger}a and Figure~\ref{fig:trigger}b, respectively. The detection efficiency can be estimated as the fraction of the area below the threshold, which is 99$\%$ for both MDA and MDB detectors when the threshold is chosen to be 40 mV. The measured occurrence rate is 1/4.74 s as shown in Figure~\ref{fig:trigger}c (a replot of Figure~4c in the main text), corresponding to the muon flux of 0.506$\pm$0.007 min$^{-1}$cm$^{-2}$ when the area of 5$\times$5 cm$^2$ of the detector is considered.

Based on a modified Gaisser formula \cite{guan_parametrization_2015}, the expected muon flux of a plane detector placed horizontally at the sea level is 0.834 min$^{-1}$cm$^{-2}$. Considering the detection efficiency of 99$\%$ for both detectors would lead to a modified flux of 0.817 min$^{-1}$cm$^{-2}$. Figures~\ref{fig:trigger}d and e show the distribution of measured voltage signals of the muon detector in the laboratory and on the top of the laboratory building. Fitting the data, we obtain the occurrence rates of 1/4.79 s and 1/3.19 s, corresponding to the muon flux of 0.501$\pm$0.008 min$^{-1}$cm$^{-2}$ and 0.752$\pm$0.007 min$^{-1}$cm$^{-2}$, respectively. These results indicate a reduction of 33$\%$ in the muon flux due to the shielding of laboratory environments, and there is a difference between the theoretical and measured muon flux of 0.817 and 0.752 min$^{-1}$cm$^{-2}$ at room temperature.

In the main text, the rate of muon-induced QP bursts is measured to be (1/67 s)/(1.5$\times$1.5 cm$^2$) $\times$ 60 $\approx$ 0.40 min$^{-1}$cm$^{-2}$, considering the entire area of the qubit chip of 1.5$\times$1.5 cm$^2$. The value is reasonable from the following estimation. If we consider the area of 4.2$\times$7.8 mm$^2$ where the 31 selected qubits locate, and $3.5\sim 4.0$~mm distance over which the ``hot point" caused by high-energy radiation decays~\cite{martinez2019measurements}, an effective area between 1.12$\times$1.48 and 1.28$\times$1.5 cm$^2$ will result, leading to the muon-induced QP burst rate between 0.54 and 0.47 min$^{-1}$cm$^{-2}$, which fairly matches the flux of 0.506 min$^{-1}$cm$^{-2}$ measured by the muon detector.

\section{Estimation of particle detection threshold}

In the detection of cosmic-ray and dark matter particles that penetrate the refrigerator, the sample box, the substrate, and the superconducting films, most initial energy affecting QP tunneling is deposited in the substrate. Here we estimate the particle detection threshold based on the available experimental data.

Google's experiment shows that particles can be detected using superconducting qubit array and the bit-flip measurement when the energy deposited in the substrate is $\sim$~100 keV~\cite{mcewen_resolving_2022}. The deposited energy can be significantly lower for an effective detection from the following considerations. First of all, as discussed in the main text, the probability of a charge-parity jump induced by QP tunneling is much higher than the probability of bit flip. In our experiment, the ratio is about 18. However, the ratio can vary over a wider range from 20 to 80 by tuning the qubit parameters~\cite{catelani2011relaxation}. Second, the QP trapping in the Al films can greatly increase the QP density near the tunneling barrier. As estimated above using Equation~(\ref{equ:qp recombianation rate}), the QP density can increase by a factor of 695 in our experiment compared to that in Google's experiment~\cite{mcewen_resolving_2022} if the background powers for qp excitation are on the same level (see section IIC). Third, both Google's experiment and ours are designed for quantum computing so only a small area of the superconducting films is relevant to qubit for detection and most area serves as ground with thermal contacts via In bumps (see Fig.~5 in the main text). Considering that our device structure is similar to that in Google’s experiment, the phonon-to-QP conversion efficiency will be greatly improved~\cite{fink2024superconducting} if we remove the ground superconducting films on the qubit chip, which will not degrade the qubit performance as verified in the previous experiments~\cite{ding2025multipurpose,bao_creating_2024}). In the present experiment, the ratio of the total qubit chip area $15\times15$ mm$^2$ to the qubit related area $\sim$~2 mm$^2$ is about 112.5.

As a result, by using the MQSCPJ method and QP trapping, and removing the ground Ta films, the detection threshold can be reduced by a factor of $C$~$\sim$50$\times$700$\times$100 down to 100 keV/$C$ = 28.6 meV. The threshold can be even lower by further improving the device performance and increasing the readout and measurement fidelities. Also, the 100-keV level may not be the lowest level for the detectable energy in Google's experiment. Finally, for particles with given energy, the Al trap volume can be tuned to have optimized QP density for improving the measurement efficiency~\cite{Booth1987}

\end{document}